\documentclass[%
 reprint,
 showkeys,
nofootinbib,
 amsmath,amssymb,
 aps,
]{revtex4-2}

\usepackage{graphicx, float, subfig,}
\usepackage{bm}

\usepackage[overload]{empheq}

\usepackage{array,tabularx,booktabs}
\renewcommand{\arraystretch}{1.3}

\begin{document}

\title{Rules, agents and order}

\author{Amalia Puente}
\affiliation{Facultad de Ciencias, Universidad de Colima,
Bernal D\'\i az del Castillo 340, Col. Villas San Sebasti\'an, 
C.P. 28045, Colima, Colima, M\'exico.}

\author{Diego Radillo-Ochoa}
\affiliation{Institut de Physique, 
École Polytechnique Fédérale de Lausanne (EPFL), Lausanne,
and 
Laboratory for X-ray Nanoscience and Technologies, 
Paul Scherrer Institut, 5232 Villigen, PSI, Switzerland.}

\author{C\'esar A. Terrero-Escalante}
\email{Author names are arranged alphabetically.
	Corresponding author: cterrero@ucol.mx}
\affiliation{Facultad de Ciencias, Universidad de Colima,
Bernal D\'\i az del Castillo 340, Col. Villas San Sebasti\'an, 
C.P. 28045, Colima, Colima, M\'exico.}

\date{\today}

\begin{abstract}
Complex systems often exhibit highly structured network topologies that reflect functional constraints. 
In this work, we investigate how,
under varying combinations of system-wide selection rules and special agents,
different classes of random processes give rise to global order, 
with a focus restricted to finite-size networks. 
Using the large-$N$ Erdős–Rényi model as a null baseline, 
we contrast purely random link-adding processes with goal-directed dynamics, 
including variants of the chip-firing model and intracellular network growth,
both driven by transport efficiency. 
Through simulations and structural probes such as \textit{k-core} decomposition and \textit{HITS} centrality, 
we show that purely stochastic processes can spontaneously generate modest functional structures, 
but that significant departures from random behavior generically require two key conditions: 
critical topological complexity and dynamic alignment between topology and functionality. 
Our results suggest that the emergence of functional architectures depends not only on the presence of selection mechanisms or specialized roles, 
but also on the network's capacity to support differentiation and feedback. 
These findings provide insight into how topology-functionality relationships emerge in natural and artificial systems 
and offer a framework for using random graph baselines to diagnose the rise of global order in evolving finite-size networks.
\end{abstract}

\keywords{complex systems; complex networks; random processes; $k$-cores; authorities; self-organization.}
\maketitle

\section{Introduction}
\label{sec:intro}

Complex systems in nature and society often display highly organized network structures 
that arise from the interaction between heterogeneous agents
under specific rules. 
Over the last century,
from microscopic to macroscopic scales,
this kind of patterns has been observed in physical systems: 
lattices of interacting spins exhibit emergent phases driven by local couplings and thermal constraints
\cite{mcCoy1973}; 
gauge field configurations self-organize under symmetry and energy minimization constraints, 
with field excitations playing the roles of heterogeneous agents
\cite{montvay1994},
and scale-free behavior of a turbulent fluid
(signaled by a power-law probability distribution of the fluid elements strength)
has been modeled using a vortical-interaction network
\cite{taira2016}.
Beyond physics,
the global order present
in complex finite-size networks 
from metabolic pathway circuits to social collaboration graphs and technological infrastructures, 
also often reflects functional pressures or constraints, 
rather than being purely the result of random connectivity
\cite{barabasi2016network}. 
A central challenge in network science is to determine the conditions under which such functional organization can emerge from decentralized, stochastic processes.

This work addresses this challenge by investigating how different classes of network evolution models give rise to 
\textit{global structural order}. 
Specifically, 
we study the emergence of $k$-cores
\cite{seidman1983, dorogovtsev2006, kong2019},
and hubs
\cite{barabasi2016network,kleinberg1999},
two topological signatures of deep connectivity and hierarchical structure,
in a variety of generative processes. 
These include both purely random models and functionally guided ones, 
distinguished by the presence or absence of 
\emph{global selection rules},
which accept or reject changes based on system-level goals
(common examples in physical systems include the minimization of potential energy or the maximization of entropy), 
and \emph{special agents}, 
which play dedicated roles in the network's function
(for instance, 
predators in ecological systems or control nodes in engineered networks). 
The large-$N$ Erdős–Rényi random graph
\cite{erdos1960}
serves as a baseline representing the class 
-\textbf{no global selection rule, no special agents}-. 
Against this baseline, 
we contrast finite-size models belonging to all the classes 
generated by combining the presence and absence of those mechanisms.

Our aim is to determine under what conditions these network evolution processes depart from the random Erdős–Rényi behavior 
and produce a functionally meaningful structure. 
By using $k$-core formation, degree and HITS based hub detection as structural probes, 
we reveal when and how goal-directed dynamics influence the emergence of global order. 

In the next section, we describe the network-based random processes considered in this study.
In Sec.~\ref{sec:class}
we define the classes of models in terms of the presence of global selection rules and special agents,
and classify the processes introduced in Sec.~\ref{sec:models}.
Then,
in Sec.~\ref{sec:order},
we outline the analytical and computational methods used
to measure the degree of global order at each stage of the evolution
of those network-based complex systems.
In Sects.~\ref{sec:results} and \ref{sec:discuss}
we present the results
and perform a comparative analysis that illustrates how structure, 
function, 
and dynamics interact across different regimes of complexity.
Finally,
in the last section we present our conclusions.

\section{Network-based random processes}
\label{sec:models}

In this section, we will proceed to describe the models we used in this study.
All these systems have in common that their underlying mathematical structure is a complex network,
which evolves through the iterative addition of links chosen with some degree of randomness.
To better articulate the dynamics underlying different random network processes, 
we adopt a conceptual distinction between \textbf{graphs} and \textbf{complex networks}. 
A graph is defined as the pair $(V, E)$, 
where $V$ is a set of vertices and $E$ a set of edges. 
This representation captures the static structure of connections, 
independent of any functional behavior or temporal evolution. 
In contrast, 
we define a complex network as the pair $(G, F)$, 
where $G = (V, E)$ is its \textbf{topology} 
and $F$ is a set of rules or functions 
(the \textbf{functionality}) 
governing the evolution of attributes
(such as node states, link weights, or activity patterns),
while keeping the topology fixed
\cite{radillo2023}.

This distinction is instrumental to our analysis: 
it enables us to separate the structural properties of the network from its internal dynamics, 
and to examine how Global Selection Rules (GSR)
and the presence of Special Agents (SA) 
contribute to the emergence of structural features such as highly connected nodes and densely interlinked subgraphs.
It also allows us to reinterpret traditionally static models, 
such as the Configuration Model
\cite{bollobasi1980} 
or clustered models \cite{bhat2017}, 
as outcomes of an implicit dynamic process,
one that could be reconstructed through local interactions guided by GSRs and the behavior of SAs.
Importantly, 
the dynamical processes we study do not replicate the ensemble generation procedures of these models in a literal sense. 
Instead, 
we approach them as \emph{guided evolutionary analogues}, 
capable of reconstructing the expected statistical properties of the ensemble through iterative and rule-based mechanisms. 

As we shall see,
the impact of GSR and SA can be confined to the topology of the network
($G$), 
or it can extend to its functionality layer 
($F$), 
influencing how the system evolves internally. 
In particular,
SA can be present by design or emerge as result of local selection rules.
In purely topological models like Barabási-Albert Preferential Attachment
\cite{barabasi1999}, 
locally selected SA
(manifested as high-degree nodes attracting more connections)
operate exclusively at the level of $G$, 
with $F = \emptyset$. 
Likewise, 
in constrained random graphs like the Configuration Model and  clustered models, 
the influence of GSR on degree sequences is also restricted to $G$. 
However, 
more elaborate models examined later in this work incorporate meaningful dynamics overlaid on the graph. 
In such cases, 
the interplay between SA and GSR spans both $G$ and $F$, 
shaping not just how nodes connect but how activity, 
information, 
or functional influence propagates,
ultimately determining the system’s emergent order.

\subsection{Erdős–Rényi random graph model}
\label{ssec:er}

A classic example in the study of graph evolution is the Erdős–Rényi (ER) random graph 
\cite{erdos1960}, 
one of the simplest and most thoroughly analyzed models. 
In its standard formulation, 
a graph is constructed with $ N $ vertices, 
where each pair of vertices is independently connected with probability $ p $. 
Alternatively, 
the process
can start with graph $G = (V, \emptyset)$ 
(with $|V|=N$ number of vertices in $V$), 
and at each iteration
to add a single new edge to the current graph. 
Each new edge is chosen uniformly and independently
out of all still available edges.
In both variants, 
the result is a purely random process, 
lacking constraints or preferences in link placement.

The degree $k$ of a vertex,
the number of edges connected to it,
follows a Poisson distribution in the large-$N$ limit, 
with the average degree given by $\langle k \rangle = pN $. 
Despite the stochastic nature of this process,
as $p$ increases, 
the network topology transitions from a collection of disconnected nodes to a structure with increasingly complex connectivity.

The ER model does not incorporate either SA or GSR; 
all nodes and links are treated uniformly, 
and no additional criteria guide the formation of the network.
It also lacks functionality.
As a consequence, 
the model lacks efficiency on generating highly connected nodes or robustly organized subgraphs. 

\subsection{Barabási–Albert Preferential Attachment model}
\label{ssec:pa}

The Barabási–Albert Preferential Attachment model 
\cite{barabasi1999} 
was originally proposed to explain the spontaneous emergence of hubs and the power-law degree distributions observed in many empirical networks. 
In the classical formulation, 
the network grows by the sequential addition of new nodes, 
each of which forms links preferentially to existing nodes with higher degree.

To enable a meaningful comparison with classical random graph models such as the ER process, 
we implement a modified version of 
the Preferential Attachment model (PA) 
in which the number of nodes is fixed and the network evolves solely through the progressive addition of links. 
This contrasts with the traditional Barabási–Albert process, 
where both the number of nodes and links increase in tandem, 
typically maintaining a constant average degree. 
In our implementation, 
links are added iteratively, 
with each new link
more likely to connect to a node with higher degree,
thereby preserving the mechanism of preferential attachment while allowing the average degree to increase, 
as in the ER model.

This adaptation permits direct examination of the emergence of structural organization under comparable conditions of network density. 
In this setup, SA emerge naturally: 
nodes with higher degree accumulate links at a faster rate, 
disproportionately shaping the evolving topology. 
However, the model does not employ any system-wide selection rule; 
all changes are driven by a \textit{local selection rule}, 
probabilistic interactions without global constraints or optimization.

Like the ER model, 
the PA model is devoid of functionality,
but it promotes early hub formation and a more gradual accumulation of densely connected regions.
This contrast underscores the distinct role that SA can play in guiding topological organization, 
even in the absence of explicit GSR.

\subsection{Achlioptas process with product rule}
\label{ssec:achlioptas}

The Achlioptas process 
\footnote{Suggested by Dimitris Achlioptas at a Fields Institute workshop in 2000}
was introduced as a variation of the classical ER model 
to explore how global connectivity might be delayed or reshaped by modifying the edge selection mechanism. 
Rather than adding edges blindly, 
the process uses a rule-based strategy to select edges that inhibit the rapid coalescence of large components. 
The most studied version is the \textbf{product rule}, 
which at each iteration considers two candidate edges 
and selects the one that minimizes the product of the sizes of the connected components that the edge would merge.

In our implementation of the Achlioptas process, 
we follow a fixed-vertex framework to maintain comparability with other link-driven evolution models such as ER 
and PA. 
We begin with an empty graph of $N$ isolated vertices and progressively add one edge at a time, 
using the product rule to guide the choice. 
At each iteration, 
two candidate edges $(u_1,v_1)$ and $(u_2,v_2)$ are selected uniformly at random (avoiding self-loops and duplicates). 
For each candidate edge, 
we determine the sizes of the connected components to which $u_i$ and $v_i$ belong, 
and compute the product of these sizes. 
The edge with the smaller product is selected and added to the graph; 
ties are broken arbitrarily.

This rule systematically suppresses the early formation of large components 
by favoring edges that connect smaller structures. 
Unlike the ER model, 
where edges are added without bias and a giant component emerges smoothly at a well-defined threshold, 
the Achlioptas Process with the Product Rule 
(APPR)
can induce an abrupt, 
seemingly discontinuous percolation transition
\cite{achlioptas2009}. 
Although subsequent studies have clarified that the transition remains continuous in the thermodynamic limit 
\cite{riordan2012}, 
the delayed and abrupt emergence of large-scale connectivity remains a hallmark of this model.

Since no specific nodes are favored, no SA are present,
but link adding decisions are made globally based on the current component structure, 
so a GSR guides the evolution. 
The product rule thus provides a paradigmatic example 
of how global structural properties can be influenced without invoking node-level heterogeneity
and in the absence of network functionality.

\subsection{Other Achlioptas-like Processes}
\label{ssec:crp}

A broad class of generative models,
including the Configuration Model (CM) 
and clustered models such as the $(c, q)$-model 
\cite{bhat2017},
are typically defined through static procedures that generate random networks subject to prescribed structural constraints, 
most notably specific degree distributions and clustering levels. 
These models are not inherently dynamical: 
they define an ensemble of graphs based on global constraints but do not describe how such graphs might emerge through an evolutionary process.

For comparative purposes, 
however
(particularly in assessing the influence of GSR and the emergence of structural features)
we found
fruitful to reinterpret these static models as the outcome of an underlying constrained random process. 
In our view, 
the network evolves through iterative link additions, 
with the dynamics guided by global constraints embedded in a system-wide fitness criterion.

Let $ \mathcal{G}_{\text{target}} $ denote the ensemble of graphs satisfying the desired constraints,
e.g., 
fixed degree distribution in the CM or target average degree and clustering coefficient in the $(c, q)$-model. 
We define a constrained random process as follows:
\begin{enumerate}
    \item At each time step $ t $, a set of candidate graphs $ \mathcal{C}_t $ is generated by adding an edge to the current graph $ G_{t-1} $, 
    drawn randomly or according to a heuristic.
    \item A fitness function $ P_{\text{fit}}(G) $ is evaluated for each candidate $ G \in \mathcal{C}_t $, 
    estimating its compatibility with $ \mathcal{G}_{\text{target}} $.
    \item The candidate with the highest fitness score is selected to form the next graph $ G_t $.
\end{enumerate}

Assuming ergodicity in the graph space and reachability of the target ensemble, 
this process approximates a biased random walk over the graph space, 
increasingly favoring configurations consistent with the imposed constraints. 
By encoding the relevant constraints into $ P_{\text{fit}}(G) $, 
the procedure indirectly respects the maximum entropy principle that underlies the definition of many static models, 
including CM and clustered variants.
This perspective enables a unified reinterpretation of all these models 
as instances of Achlioptas-like processes. 
At each time step, 
a small set of candidate edges is generated, 
and the one that best aligns the evolving graph with a predefined structural ensemble,
characterized by constraints such as degree distribution or clustering,
is selected. 
The selection rule operates at a global level, 
favoring edge additions that improve a fitness function measuring proximity to the target ensemble. 
These models therefore rely solely on GSR, 
without invoking SA or vertex-level biases. 
As the process unfolds, 
structural observables progressively converge toward their prescribed values, 
and the resulting graphs reflect the statistical properties of the intended ensemble through a constrained, 
edge-wise growth dynamics.

To widen the scope of the comparison to be presented in Sec.~\ref{sec:discuss},  
we derived the data for these models from the results presented in Ref.~\cite{bhat2017},
which were obtained by applying the $k$-core peeling algorithm to static graph configurations.

\subsection{Network Representation of the Jamming Process}
\label{ssec:jamming}

The jamming process describes the transition from a fluid-like to a solid-like state in disordered systems 
such as foams, emulsions, granular media, and glasses 
\cite{liu2001}. 
In a network-theoretic representation 
\cite{morone2019}, 
the system is modeled as an underlying graph $ G = (V, E) $, 
where each vertex $ v \in V $ corresponds to a particle (e.g., a soft sphere), 
and an edge $ e = (v_i, v_j) \in E $ is formed if the associated particles are in physical contact,
typically through overlap or the transmission of mechanical force.
The corresponding network evolves through the gradual addition of links as the system is compressed or locally rearranged. 
At each step, 
link formation is governed by physical constraints such as force balance and minimal energy configurations, 
which serve as GSR guiding the next allowed configurations. 
These selection rules are system-wide, 
since they impose collective constraints
on the entire configuration. 
The process proceeds until the system reaches a \emph{jammed state}, 
marked by the formation of a globally rigid structure.

While this model does not involve explicit SA, 
the influence of GSR is paramount: 
the evolution is constrained by non-local mechanical rules rather than purely stochastic link additions. 
The resulting topological order
(including the appearance of rigid or percolating substructures)
can be analyzed in terms of the formation of highly connected regions or through the emergence of dense subgraphs under increasing connectivity
\cite{morone2019}. 
This abstraction enables meaningful comparisons with other link-driven models, 
especially in understanding how system-wide selection rules shape structural transitions.

\subsection{Modified Chip-Firing Model}
\label{ssec:mcfm}

The Chip-Firing Model (CFM) 
is a prototypical example of \emph{self-organized criticality}, 
demonstrating how complex, scale-invariant behavior
(such as cascades or avalanches)
can emerge from simple local rules. 
It has been introduced into the literature a number of times from various communities,
because has deep connections with several fields, 
including Combinatorics 
(e.g., spanning trees, recurrent configurations), 
Algebra 
(e.g., the sandpile or critical group of a graph), 
and Statistical Physics.
Its commonly accepted origin is as a straightforward version of the abelian sandpile model
\cite{bak1987}.

The classical CFM is a discrete dynamical system defined on a network 
with topology given by a finite \textit{directed} graph $ G = (V, E) $. 
Each node $ v \in V $ holds as attribute a non-negative integer $ c(v) $, 
representing a generic transferable resource 
(e.g., sand grains, tokens, or energy units). 
According with its functionality rules,
at each discrete time step, 
a node $ v \in V $ is \emph{active} if $ c(v) \geq \deg(v) $. 
When such a node fires, 
it distributes one chip to each of its neighbors and reduces its own count by its degree:
$c(v) \mapsto c(v) - \deg(v)$
and 
$\forall u \in \mathrm{Nbr}(v),\ c(u) \mapsto c(u) + 1$.
Firing can occur sequentially or in parallel across multiple active nodes. 
A configuration is said to be stable if no nodes are active.

A key property of the CFM is that it is \emph{Abelian}: 
from any initial state,
the final stable configuration is independent of the order in which firings occur. 
To ensure eventual stabilization, 
SA are typically introduced
as one or more \emph{sinks},
nodes that absorb chips but never fire.

\vspace{1em}
\noindent
\textbf{Modified Chip-Firing Model.}
We implemented for this study a generalization of the classical CFM in which both external driving and structural evolution are introduced. 
This Modified Chip-Firing Model (MCFM) 
operates on a dynamically evolving directed network with graph $G_t = (V, E_t)$, 
where $t$ stands for discrete time steps. 
The fixed node set $ V $ is partitioned into three categories:
\begin{itemize}
    \item \textbf{Sink nodes} $ (V_{\text{sink}}) $: absorb chips but never fire.
    \item \textbf{Source nodes} $ (V_{\text{source}}) $: at each step receive chips only from an external reservoir.
    \item \textbf{Regular nodes} $ (V_{\text{reg}} = V_t \setminus (V_{\text{sink}} \cup V_{\text{source}})) $: follow the standard chip-firing dynamics.
\end{itemize}
At the start of a simulation,
regular nodes are initialized with a fixed random number of chips,
and those values are saved for further reference.
Then,
at the beginning of each iteration of the process:
\begin{itemize}
    \item Sink nodes are initialized with zero chips.
    \item Source nodes are initialized with enough chips to guarantee they are active; 
    their chip count is set to match their current degree  $ c(v) \geq k_v $.
    \item Regular nodes are reset to the initial values.    
\end{itemize}
Network evolution is governed by a constrained random process. At each iteration:
\begin{enumerate}
    \item A candidate edge is proposed and added to the current graph $ G_t $, 
    producing a modified graph $ G_{t+1} $.
    \item A full firing process is run using the standard rules, and the system stabilizes.
    \item The average number of chips absorbed by the sink nodes is measured.
    \item If this average exceeds that of the previous iteration, the change is accepted; otherwise, the graph reverts to $ G_t $.
\end{enumerate}

This selection rule introduces a feedback mechanism 
that guides structural evolution toward configurations that enhance chip absorption by the sinks,
representing a kind of system-level optimization. 
The presence of source and sink nodes models an open system with inputs and outputs, 
given their non-standard behavior
they function as SA,
while the evolution rule defines a GSR.

We examine two variants of this model:
\begin{itemize}
    \item An \textbf{unconstrained} version without GSR: all randomly chosen edges are accepted regardless of their impact on sink absorption.
    \item A \textbf{constrained} version with GSR: only proposed edges that improve sink absorption are accepted.
\end{itemize}
This distinction allows us to isolate and study the structural consequences of incorporating global selection mechanisms 
in dynamical network models with the same functionality.

\subsection{Evolutionary Model of a Single-Cell Organism Based on Abstract Chemistry}
\label{ssec:seoe}

In several previous studies 
\cite{ibarra-junquera2022,radillo2023}
we have considered a constrained random process inspired by Kauffman's models of abstract chemical reaction networks
\cite{kauffman1969,kauffman1969a}, 
intended to simulate the evolutionary dynamics of a single-cell organism. 
The abstract cell consists of a set of $ N $ \textit{chemical species} 
$ \{S_1, S_2, \dots, S_N\} $, 
whose interactions are governed by a system of ordinary differential equations (ODEs). 
The topology of the underlying network is a finite \textit{oriented} graph $ G = (V, E) $, 
where each vertex $ v \in V $ represents a chemical species, 
and each directed edge $ (v_i, v_j) \in E $ represents an \textit{irreversible catalytic reaction} from species $ S_i $ to species $ S_j $.
As part of its functionality,
each species $ S_i $ has a time-dependent concentration $ x_i(t) $, 
which evolves according to a system of coupled ODEs:
\[
\frac{dx_i}{dt} = f_i(x_1, x_2, \dots, x_N)
\]
where the function $ f_i $ models the net production and degradation of species $ S_i $, 
influenced by other species via catalytic and reactive effects,
as well as the process of nutrient transport from the environment.

Within the network, we distinguish three functional classes of species:
\begin{itemize}
    \item \textbf{Nutrients:} external and internal species whose concentrations are exhausted by the production of other species, 
    and restored by diffusion and environmental gradients.
    \item \textbf{Nutrient carriers:} internal species that facilitate the influx of nutrients; their concentrations modulate nutrient uptake rates and are regulated by the internal network.
    \item \textbf{Catalysts:} regular internal species that influence the production rates of others by enhancing reaction efficiencies.
\end{itemize}

Evolutionary dynamics are simulated through a \textit{mutation–selection process} over generations:
\begin{enumerate}
    \item Begin with a population of $ m $ \emph{mother networks}.
    \item Each mother network is mutated by randomly adding a new chemically plausible link, yielding a population of \emph{daughter networks}.
    \item For each daughter network, the associated system of ODEs is numerically integrated until a specified \emph{duplication condition} is met. 
    This condition is defined as the doubling of the initial cell volume, which is computed as a functional of the concentrations.
    \item The $ m $ daughter networks that reach the duplication threshold in the shortest time are selected to form the next generation of mothers.
\end{enumerate}

This Intracellular Network Evolution Process (INEP)
couples \emph{plausibility-constrained random mutations} (link additions) 
with an externally imposed GSR based on replication efficiency.
Nutrients and nutrient carriers act as SA that significantly modulate system behavior through their regulatory roles in nutrient intake and reaction control.
Despite its abstract, artificial nature,
versions of the INEP have been successful in replicating properties of real organisms like differentiation and
pluripotency \cite{furusawa2009}, 
power-law chemical abundance \cite{furusawa2003,radillo2023}, 
the reduction in the dimensionality of the phenotypic space changes due to environmental perturbations \cite{sato2020}, 
the fact that the ‘outward’ central metabolites correspond mainly to building-block molecules, 
the rise of the small-world structure, 
the existence of a small number of key metabolites exhibiting the highest degree of connectivity
\cite{ibarra-junquera2022},
as well as the ingrained
but sudden change
from a stagnant phase to exponential growth
as the evolutionary asymptotic state of a single-cell organism
\cite{radillo2023}.

\section{Classification of Network-Based Random Processes} \label{sec:class}

Having introduced a diverse set of network-based random processes in the preceding section,
we now turn to a unifying classification scheme.
This framework is based on two independent features that shape the evolution of these systems:
the presence of a \textbf{global selection rule}, 
which evaluates and favors certain configurations,
and the presence of \textbf{special agents}, 
nodes or components with distinct roles or properties that influence the system's behavior.
These two binary features yield four distinct classes of network-based dynamics:
\begin{itemize} 
\item \textbf{No Global Selection Rule, No Special Agents (nGSRnSA)} \\ 
In this baseline class, 
structural changes occur through uniform randomness or homogeneous local rules.
There is no global evaluation to guide evolution, 
and no node plays a privileged role.
If global structure emerges, 
it does so purely as a collective effect of symmetric interactions.
The classical ER model belongs here.

\item \textbf{No Global Selection Rule, Special Agents (nGSRSA)} \\
Systems in this class lack explicit selection or optimization,  
but include components
(such as hubs, sources, or sinks)
that introduce structural or functional asymmetries.  
These SA influence local dynamics,
while being themselves subject to different rules for evaluation or change.  
An example is the MCFM with fixed sources and sinks but no sink optimization.

\item \textbf{Global Selection Rule, No Special Agents (GSRnSA)} \\
Here, all elements are structurally equivalent, 
but the system is guided by a GSR.  
This rule evaluates configurations against system-wide criteria 
(e.g., target degree distribution or clustering),  
favoring those that better meet these objectives.  
Typical examples include the CM and its clustered variants.  

\item \textbf{Global Selection Rule, Special Agents (GSRSA)} \\
Systems in this class combine selection mechanisms with internal heterogeneity.  
SA actively influence system behavior,
acting as entry points, regulators, or drivers,  
while global performance is continually evaluated.  
Evolution proceeds via both internal feedback and external selection.  
The INEP is a prototypical example in this class. 

\end{itemize}

This classification captures the spectrum from neutral, 
homogeneous dynamics to highly structured, 
adaptive systems.
It provides a useful map for understanding the design space of network-based random processes,
for instance those used in this study,
whose classification is summarized in table \ref{tab:class}.
\begin{table*}[t]
\centering
\renewcommand{\arraystretch}{1.5}
\begin{tabular}{>{\raggedright}p{0.2\textwidth}>{\raggedright}m{0.28\textwidth}>{\raggedright}m{0.26\textwidth}>{\raggedright\arraybackslash}m{0.08\textwidth}}
\toprule[1pt]
\textbf{Model} & \textbf{Global Selection Rule} & \textbf{Special Agents} & \textbf{Class} \\
\midrule[1pt]
ER & No  & No  & nGSRnSA \\
\midrule[0.1pt]
PA & No & \parbox[m]{0.28\textwidth}{\raggedright Emergent; nodes with higher connectivity} & nGSRSA \\
\midrule[0.1pt]
Unconstrained MCFM & No & Designed; sources and sinks & nGSRSA \\
\midrule[0.1pt]
APPR & \parbox[p]{0.28\textwidth}{\raggedright \vspace*{3pt} Smallest product of connected components sizes \vspace*{3pt}} & No & GSRnSA \\
\midrule[0.1pt]
CM & \parbox[m]{0.28\textwidth}{\raggedright \vspace*{3pt}Closest to a given degree distribution \vspace*{3pt}} & No & GSRnSA \\
\midrule[0.1pt]
Clustered ER  & \parbox[m]{0.28\textwidth}{\raggedright \vspace*{3pt} Closest to a given average degree and clustering \vspace*{3pt}} & No & GSRnSA \\
\midrule[0.1pt]
Clustered CM & \parbox[m]{0.28\textwidth}{\raggedright \vspace*{3pt} Closest to a given degree distribution and clustering \vspace*{3pt}} & No & GSRnSA \\
\midrule[0.1pt]
Jamming & \parbox[m]{0.28\textwidth}{\raggedright \vspace*{3pt} Strongest fulfillment of mechanical constraints \vspace*{3pt}} & No & GSRnSA \\
\midrule[0.1pt]
Constrained MCFM & Highest sink absorption & Designed; sources and sinks & GSRSA \\
\midrule[0.1pt]
INEP & Fastest duplication time & Designed; nutrients and carriers & GSRSA \\
\bottomrule[1pt]
\end{tabular}
\caption{Classification of models based on the presence of global selection rules and special agents.}
\label{tab:class}
\end{table*}

\section{Structural order in network-based complex systems}
\label{sec:order}

To understand the outcomes of self-organization in complex network-based systems, 
we need a way to measure the degree of \emph{structural order} that emerges from their evolution. 
Different dynamical rules and configurations can lead to networks with highly variable topologies,
some disordered and homogeneous, 
others hierarchical, clustered, or core-periphery structured. 
We focus on two key topological features as quantitative proxies for self-organization: 
\textbf{hubs} and \textbf{k-cores}.
Both represent distinct manifestations of structural centralization and connectivity that arise through non-trivial correlations and interactions.

In network theory, 
a \textbf{hub} is typically a node that has significantly more connections than average, 
acting as a central connector or bottleneck in the flow of information, resources, or influence. 
Hubs often emerge in growing networks that follow a local selection rule
(for instance, preferential attachment), 
leading to a power-law degree distribution.

We identify hubs in two complementary ways:
\begin{itemize}
    \item \textbf{By degree}: Nodes whose degree lies above a certain threshold 
    (e.g., top 1\% or those exceeding a fixed multiple of the average degree) are designated as hubs.
    \item \textbf{By HITS centrality}: The Hyperlink-Induced Topic Search (HITS) algorithm 
    \cite{kleinberg1999} 
    assigns to each node two scores: 
    \emph{authority} and \emph{hubness}. 
    Nodes with high hub scores are those that point to many authoritative nodes,
    and nodes with high authority are those that are pointed by many hubs.
    Both serve as a global indicator of topological prominence, 
    especially in directed or functionally asymmetric networks.
\end{itemize}

Another fundamental concept in graph analysis is the \textbf{k-core},
 $\mathcal{K}_k(G) \subseteq G$,
which is a maximal subgraph where every vertex has at least $ k $ connections within the subgraph 
\cite{seidman1983}. 
A large number of theoretical studies of $k$-cores has addressed fundamental questions: 
Will a $k$-core emerge as a graph evolves? 
If so, at what critical mean degree does this occur?
How large is the resulting core compared to the entire graph?
What kind of phase transition does this emergence represent?
(See Refs.~\cite{dorogovtsev2006,kong2019} for reviews of these topics.)
In some cases (e.g., classical large-$N$ ER
\cite{pittel1996}), 
analytical answers can be obtained, 
providing thresholds and scaling laws. 
But as model complexity increases,
such predictions become analytically intractable,
so they are estimated from real-world or simulations data.
The process of finding these cores,
\textit{the k-core peeling algorithm},
involves recursively removing vertices with degree less than $ k $ until no further vertices can be eliminated. 
This reveals hierarchical structures within graphs 
and plays a key role in understanding network resilience and percolation transitions.
In diverse applications
(ranging from jamming transitions in granular materials
\cite{morone2019}
to influence propagation in social systems
\cite{kong2019})
the formation of $k$-cores signals a shift from disorder to structured organization. 
Thus, $k$-core analysis offers a powerful lens into the internal cohesion and structural depth of evolving networks.

Assume $\mathcal{K}_k(G)$ is a given $k$-core of the graph $G$ then, 
in the kind of phase transition giving rise to $k$-cores, 
the order parameter is the size of the connected component, 
often expressed as the fraction of nodes it contains
\cite{newman2003}
\[
r_G \equiv \frac{|\mathcal{K}_k(G)|}{N} \, ,
\]
where $|\mathcal{K}_k(G)|$ denote the number of vertices in component $\mathcal{K}_k(G)$.
Depending on the nature of the phase transition,
below the critical average degree $\left<k\right>_k$,
$r_G$ is typically of size 
$O(log N)$ 
(for instance,
$k= 1, 2$,
second-order phase transition in ER)
or essentially zero
(for instance,
$k\geq 3$,
first-order phase transition in ER). 
Above $\left<k\right>_k$,
a component of size proportional to $N$ emerges.
In analogy with statistical physics, 
where susceptibility quantifies the system’s response to external perturbations, 
the \emph{susceptibility} 
$S(\left<k\right>)$ of a network-based random process  with respect to $k$-core formation 
can be defined
as 
\cite{newman2003}: 
\[
S(\left<k\right>) = \frac{1}{N} \sum_{i=1}^r |G_i|^2,
\]
where 
$G_1, G_2, \dots, G_r$ are the connected components of the $k$-core complement $G \setminus \mathcal{K}_k(G)$, 
and
$N$ is the total number of nodes in the network. 
It captures the typical size of regions not yet integrated into the $k$-core, 
and thus indicates how susceptible the system is to forming denser, 
globally connected structures as the external field
(modeled here by the progressive addition of links)
is increased.
Two kind of reactions can be measured,
the response to adding an undirected link
(\textit{weakly-connected susceptibility}, $S_w(\left<k\right>)$)
and to adding a directed one
(\textit{strongly-connected susceptibility}, $S_s(\left<k\right>)$).

The main difference between hubs and $k$-cores is that hubs
typically result from \textit{local selection rules}, 
like preferential attachment or similar growth mechanisms,
which focuses on individual nodes and their immediate neighborhood.
Meanwhile, 
$k$-cores arise due to \textit{global structural cohesion}, 
often analyzed in percolation-like transitions.
Despite their differences,
hubs and $k$-cores share important structural similarities.
They both
reflect non-random concentrations of connectivity,
signal the presence of global order,
and are central to the functioning and robustness of the system.

\section{Results}
\label{sec:results}

In this section, 
we present the results characterizing the structural evolution of the finite-size networks generated by various models under study. 
We begin by describing $k$-core formation and its sharpness across models using normalized measures, 
followed by an examination of hub emergence and centrality structure using HITS scores.

Table~\ref{tab:av_degree} presents the average degree $\left<k\right>_k$ at which the $k$-core first appears for $k=2$, $3$, $4$, and $5$.
\begin{table*}[t]
\centering
\renewcommand{\arraystretch}{1.5}
\begin{tabular}{>{\raggedright}m{0.06\textwidth} >{\raggedright}m{0.28\textwidth} >{\centering}m{0.12\textwidth} >{\centering}m{0.15\textwidth} >{\centering}m{0.12\textwidth} >{\centering\arraybackslash}m{0.12\textwidth}}
    \toprule
         \textbf{Row} & \textbf{Model} & $\left<k\right>_2$ & $\left<k\right>_3$ & $\left<k\right>_4$ & $\left<k\right>_5$ \\
    \midrule
         1 & \parbox[m]{0.28\textwidth}{\raggedright \vspace*{6pt} ER [\textbf{nGSRnSA}] \\ $N\to \infty$ \cite{pittel1996} \vspace*{6pt}} & 1.00 & 3.35 & 5.14 & 6.81 \\
          2 &
          \parbox[m]{0.28\textwidth}{\raggedright \vspace*{6pt} ER [\textbf{nGSRnSA}] \\ $N=10000$ [simulations] \vspace*{6pt}} 
          & $0.88 \pm 0.09$ & $3.34 \pm 0.01$ & $5.14 \pm 0.02$ & $6.79 \pm 0.02$ \\
          
          3 &
          \parbox[m]{0.28\textwidth}{\raggedright \vspace*{6pt} ER [\textbf{nGSRnSA}] \\ $N=500$ [simulations] \vspace*{6pt}}
          & $0.97 \pm 0.21$ & $3.31 \pm 0.08$ & $5.09 \pm 0.08$ & $6.73 \pm 0.08$ \\
          4 &
           \parbox[m]{0.28\textwidth}{\raggedright \vspace*{6pt} PA [\textbf{nGSRSA}] \\ $N=10000$ [simulations] \vspace*{6pt}} 
          & $1.01 \pm 0.01$ & $2.70 \pm 0.03$ & $4.41 \pm 0.03$ & $6.13 \pm 0.03$ \\
          5 &
           \parbox[m]{0.28\textwidth}{\raggedright \vspace*{6pt} PA [\textbf{nGSRSA}] \\ $N=500$ [simulations] \vspace*{6pt}} 
          & $1.03 \pm 0.04$ & $2.64 \pm 0.17$ & $4.35 \pm 0.12$ & $6.04 \pm 0.14$ \\       
          6 &
           \parbox[m]{0.28\textwidth}{\raggedright \vspace*{6pt} Unconstrained MCFM [\textbf{nGSRSA}] \\ $N=500$ [simulations] \vspace*{6pt}} 
          & $1.07 \pm 0.08$ & $3.29 \pm 0.08$ & $5.09 \pm 0.08$ & $6.71 \pm 0.09$ \\  
          7 &
           \parbox[m]{0.28\textwidth}{\raggedright \vspace*{6pt} APPR [\textbf{GSRnSA}] \\ $N=10000$ [simulations] \vspace*{6pt}} 
           & $1.61 \pm 0.17$ & $3.56 \pm 0.01$ & $5.24 \pm 0.01$ & $6.83 \pm 0.02$ \\      
         8 &
          \parbox[m]{0.28\textwidth}{\raggedright \vspace*{6pt} APPR [\textbf{GSRnSA}]\\$N=500$ [simulations] \vspace*{6pt}} 
         & $1.60 \pm 0.19$ & $3.53 \pm 0.05$ & $5.20 \pm 0.07$ & $6.76 \pm 0.08$ \\       
         9 &
         \parbox[m]{0.28\textwidth}{\raggedright \vspace*{6pt} CM [\textbf{GSRnSA}] \\ $N\to \infty$ \cite{bhat2017} \vspace*{6pt}} 
         & $---$ & For $q\in [0.01, 0.5]$, \newline $\left<k\right>_3\in( 3.30, 2.38)$ & $---$ & $---$ \\                      
         10 &
         \parbox[m]{0.28\textwidth}{\raggedright \vspace*{6pt} Clustered ER [\textbf{GSRnSA}] \\ $N=10^6, 10^7$, $q=(0.01, 0.5$) \cite{bhat2017} \vspace*{6pt}}
         & $---$ & For $q\in[0.01, 0.5]$, \newline $\left<k\right>_3\in( 3.30, 1.50)$ & $---$ & $---$ \\               
         11 &
         \parbox[m]{0.28\textwidth}{\raggedright \vspace*{6pt} Clustered CM [\textbf{GSRnSA}] \\ $N=10^6, 10^7$, $q=(0.01, 0.5$) \cite{bhat2017} \vspace*{6pt}}
         & $---$ & For $q\in[0.01, 0.5]$, \newline $\left<k\right>_3\in( 3.30, 0.78)$ & $---$ & $---$ \\
         12 &
         \parbox[m]{0.28\textwidth}{\raggedright \vspace*{6pt} Jamming [\textbf{GSRnSA}] \\ $N=2000$ \cite{morone2019} \vspace*{6pt}}
         & $\ngeq 2.00$ & $\gtrsim 3.35$ & $\gtrsim 5.14$ & $---$ \\         
         13 &
         \parbox[m]{0.28\textwidth}{\raggedright \vspace*{6pt} Constrained MCFM [\textbf{GSRSA}] \\ $N=500$ [simulations] \vspace*{6pt}}
         & $1.03 \pm 0.04$ & $2.46 \pm 0.18$ & $4.32 \pm 0.18$ & $6.38 \pm 0.17$ \\
         14 &
         \parbox[m]{0.28\textwidth}{\raggedright \vspace*{6pt} INEP [\textbf{GSRSA}] \\ $N=500$ [simulations] \vspace*{6pt}} 
         & $\ngeq 2.00$ & $2.48 \pm 0.12$ & $3.27 \pm 0.19$ & $4.20 \pm 0.21$ \\
    \bottomrule
\end{tabular}
    \caption{Average degree at which the $k$-core first appears, 
    $\left<k\right>_k$, 
    for each model. 
    Model class is indicated in brackets. 
    Results are based on simulations or literature sources, as noted. 
    "---" indicates unavailable data.}
    \label{tab:av_degree}
\end{table*}

Where available, we report both analytical results and numerical simulations, 
indicating the network size $N$ used in each case. 
For models taken from previous studies, 
the relevant literature source is cited. 
Entries marked with "---" reflect cases where no reliable data were available.
Some simulations, 
notably for the Jamming and INEP models, 
exhibited a nontrivial 2-core containing more than 50\% of the nodes at the initial average degree $\left<k\right> = 2$. 
This suggests that the actual threshold for 2-core emergence lies below this value. 
These cases are indicated with the symbol $\ngeq 2.00$.
The rise of the $k$-cores in CM and clustered models
depend on the probability $q$,
so for those cases this probability and the corresponding $\left<k\right>_3$
are shown as intervals.

As a reference,
the first row in Table~\ref{tab:av_degree} shows the analytical thresholds for ER graphs in the infinite-size limit, 
as derived by Pittel, Spencer, and Wormald~\cite{pittel1996}. 
While the ER model belongs to the nGSRnSA class, 
it still exhibits structural self-organization. 
To explore finite-size effects, 
we simulated the ER model for $ N = 10000 $ and $ N = 500 $, 
generating 100 realizations for each case. 
We tracked the emergence of each $ k $-core and computed the corresponding average degree at which it first appears. 
These results (rows 2 and 3 of the table) show excellent agreement with the theoretical predictions, 
even for moderately sized systems.

We adopted the same procedure for the PA (nGSRSA) 
and APPR (GSRnSA) models 
(rows 4 and 5, and 7 and 8, respectively), 
revealing that in PA $k$-cores appear at lower average degrees than in the ER model, 
reflecting the influence of degree heterogeneity,
while in APPR the rise of the $2$-core is delayed as consequence of the product rule.
Thresholds for the Jamming model (row 12, GSRnSA) were extracted from Ref.~\cite{morone2019}, 
while the previous three rows include results for clustered models (GSRnSA)
derived from Ref.~\cite{bhat2017}, 
where each model introduces increasing structural constraints.
For the MCFM, 
we distinguish between cases without selection rules 
(unconstrained, nGSRSA) 
and with selection 
(constrained, GSRSA), 
reported in rows 6 and 13. 
We performed 50 simulations per case on networks with $N = 500$ nodes. 
Nodes 0–3 were designated as sources, and nodes 4–7 as sinks.
Wrapping up with the models studied here,
the results for the INEP 
(row 14, GSRSA) 
are based on 30 simulations, 
each one involving 10 mother and 20000 daughter networks per iteration, 
all with $N=500$ nodes. 
Here, nodes 0–3 acted as nutrients and nodes 4–7 as nutrient carriers.

\begin{figure} [H]
    \centering
    \subfloat[Order parameter, $r_G$]{
   \label{uMCFMkcores}
    \includegraphics[width=0.47\textwidth,height=6.5cm]{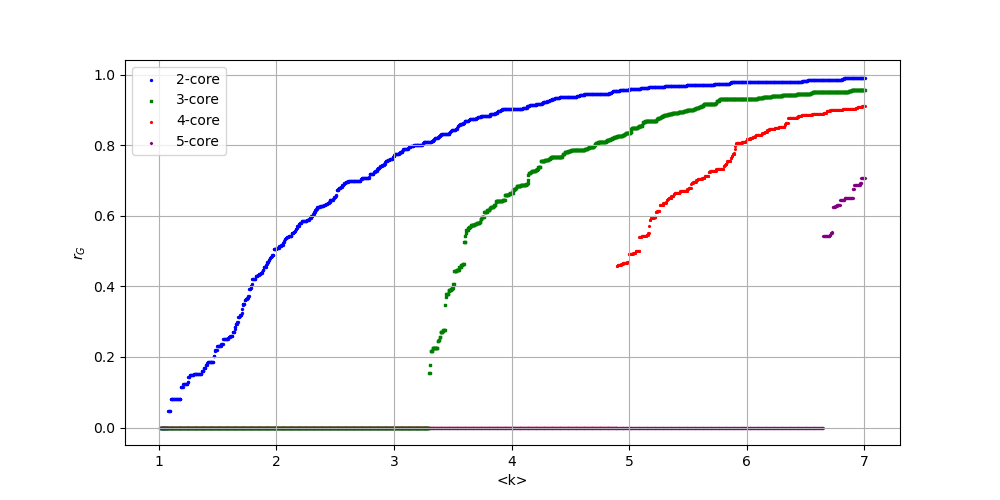}}\\
  \subfloat[Weakly-Connected Susceptibility, $S_w(\left<k\right>)$]{
   \label{uMCFMS}
    \includegraphics[width=0.47\textwidth,height=6.5cm]{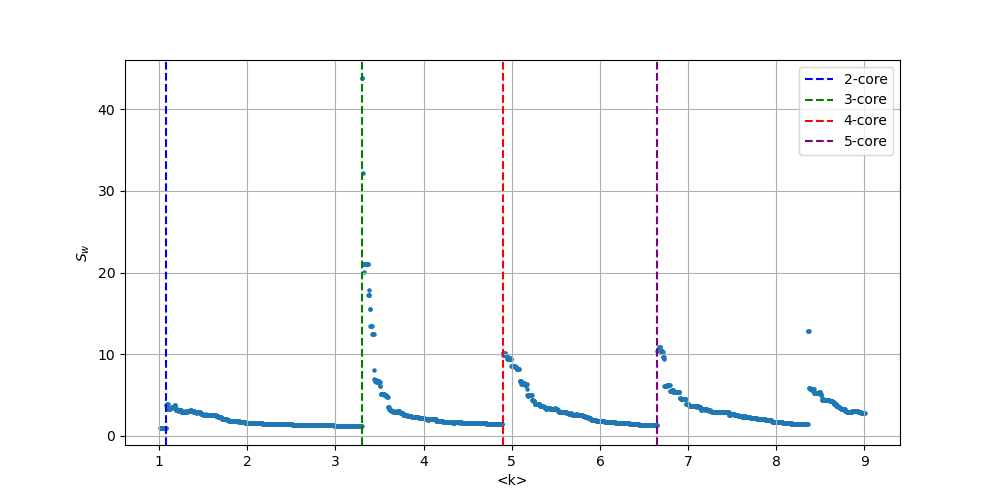}}
    \\\subfloat[Strongly-Connected Susceptibility, $S_s(\left<k\right>)$]{
   \label{2core}
    \includegraphics[width=0.47\textwidth,height=6.5cm]{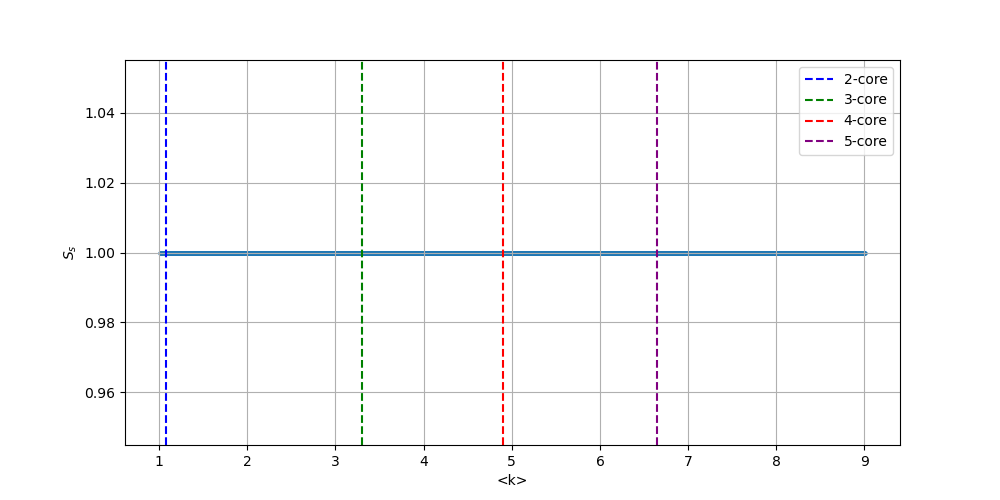}}\\
    \caption{$K-$core evolution of unconstrained MCFM as function of $\langle k\rangle$. 
    The 2-core starts at $\langle k\rangle \approx1.07$, 
    the 3-core emerges at $\langle k\rangle \approx3.29$, 
    the 4-core emerges at $\langle k\rangle \approx5.09$, 
    finally the 5-core emerges at $\langle k\rangle \approx6.71$.}
    \label{fig:uMCFMkcores}
\end{figure}

\begin{figure}[H]
    \centering
    \subfloat[Order parameter, $r_G$]{
   \label{MCFMkcores}
    \includegraphics[width=0.47\textwidth,height=6.5cm]{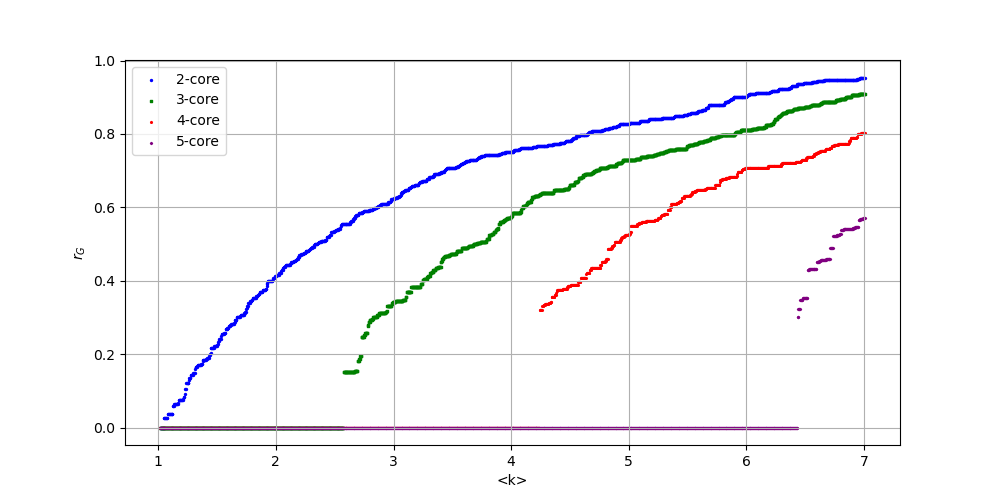}}\\
  \subfloat[Weakly-Connected Susceptibility, $S_w(\left<k\right>)$]{
   \label{MCFMS}
    \includegraphics[width=0.47\textwidth,height=6.5cm]{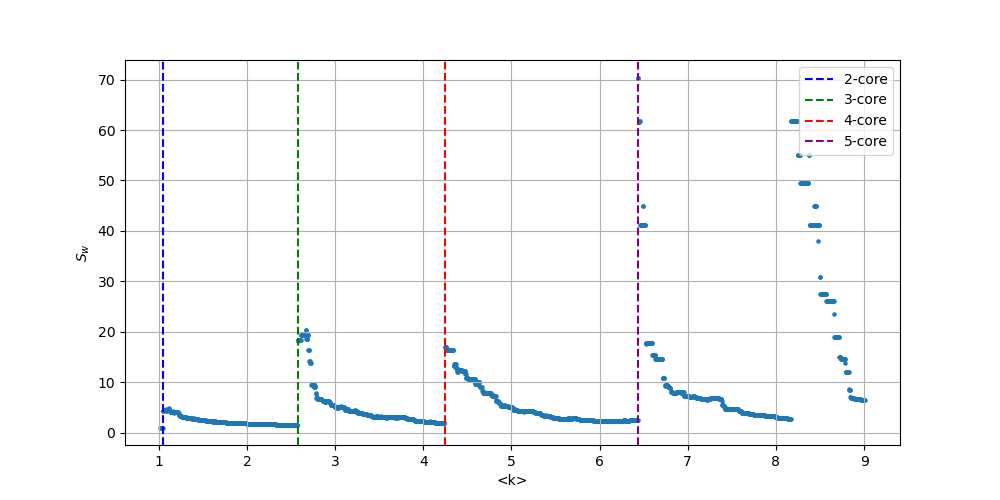}}\\
    \subfloat[Strongly-Connected Susceptibility, $S_s(\left<k\right>)$]{
   \label{2core}
    \includegraphics[width=0.47\textwidth,height=6.5cm]{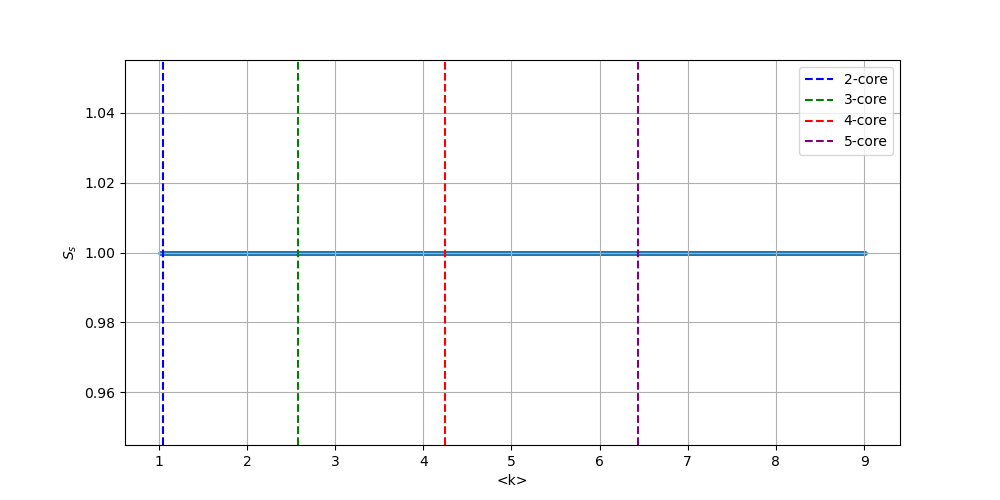}}\\
    \caption{$K-$core evolution of constrained MCFM as function of $\langle k\rangle$. 
    The 2-core starts at $\langle k\rangle \approx1.03$, 
    the 3-core emerges at $\langle k\rangle \approx2.46$, 
    the 4-core emerges at $\langle k\rangle \approx4.32$, 
    finally the 5-core emerges at $\langle k\rangle \approx6.38$.}
    \label{fig:cMCFMkcores}
\end{figure}

\begin{figure} [H]
    \centering
    \subfloat[Order parameter, $r_G$]{
   \label{INEPkcores}
    \includegraphics[width=0.47\textwidth,height=6.5cm]{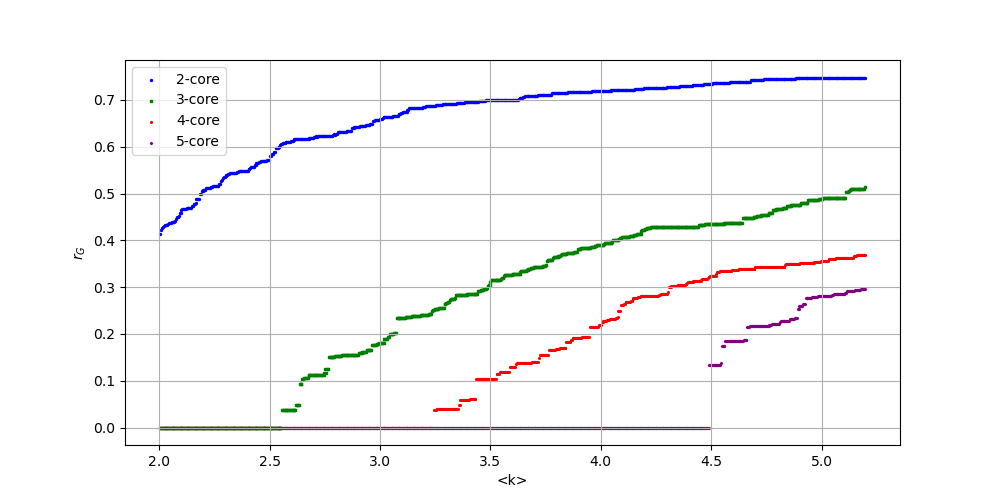}}\\
  \subfloat[Weakly-Connected Susceptibility, $S_w(\left<k\right>)$]{
   \label{INEPS}
    \includegraphics[width=0.47\textwidth,height=6.5cm]{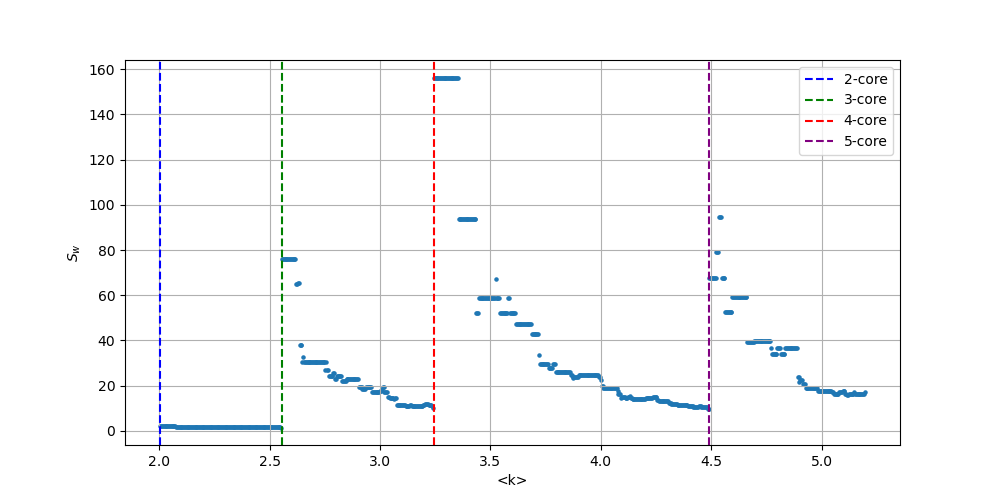}}\\
    \subfloat[Strongly-Connected Susceptibility, $S_s(\left<k\right>)$]{
   \label{2core}
    \includegraphics[width=0.47\textwidth,height=6.5cm]{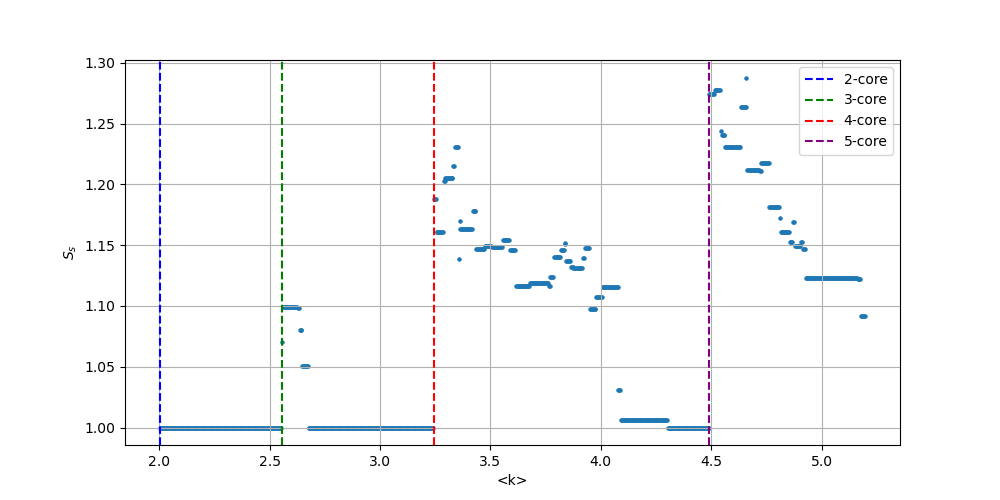}}\\
    \caption{$K-$core evolution of INEP as function of $\langle k\rangle$. 
    The 2-core starts at $\langle k\rangle \ll 2$, 
    the 3-core emerges at $\langle k\rangle \approx2.48$, 
    the 4-core emerges at $\langle k\rangle \approx3.27$, 
    finally the 5-core emerges at $\langle k\rangle \approx4.20$.}
    \label{fig:cellkcores}
\end{figure}

Examples of the evolution of the order parameter $r_G$
and the behavior of the susceptibilities $S_w(\left<k\right>)$
and
$S_s(\left<k\right>)$
in both,
unconstrained and constrained MCFM, 
are presented in figures \ref{fig:uMCFMkcores} and \ref{fig:cMCFMkcores}. 
A corresponding example for the INEP is shown in Fig.~\ref{fig:cellkcores}.

To further characterize the dynamics of $k$-core formation beyond the 3-core, 
we define two normalized measures: $\delta_{4,3}$ and $\delta_{5,4}$. 
These quantities evaluate how the interval between successive $k$-cores in a given model 
compares to that in the ER reference model.
Specifically, we define:
\begin{eqnarray}
\delta_{4,3} &=& \frac{ \left( \left<k\right>_4 - \left<k\right>_3 \right)_{\text{model}} }{ \left( \left<k\right>_4 - \left<k\right>_3 \right)_{\text{ER}} },
\quad \\
\delta_{5,4} &=& \frac{ \left( \left<k\right>_5 - \left<k\right>_4 \right)_{\text{model}} }{ \left( \left<k\right>_5 - \left<k\right>_4 \right)_{\text{ER}} }.
\end{eqnarray}
Here, $\delta_{4,3}$ measures the relative width of the transition from 3-core to 4-core, while $\delta_{5,4}$ compares the 4-core to 5-core transition. 
A value of $\delta \approx 1$ indicates behavior similar to the large-$N$ ER benchmark. 
Values of $\delta < 1$ indicate an accelerated transition between successive $k$-cores relative to ER, 
while $\delta > 1$ signals a delayed transition. 
Table~\ref{tab:delta_kcores} reports these normalized values for each model. 
For models where the $k$-core emergence is less sharply defined or data are unavailable, 
we use appropriate annotations as before.
\begin{table}[ht]
\centering
\begin{tabular}{p{5cm} p{1.5cm}<{\centering} p{1.5cm}<{\centering}}
\toprule
\textbf{Model} & \bm{$\delta_{4,3}$} & \bm{$\delta_{5,4}$} \\
\midrule
         ER (N=10000) & $0.95$ & $0.99$ \\
         ER (N=500) & $0.99$ & $0.98$ \\
         PA (N=10000) & $0.95$ & $1.03$ \\
         PA (N=500) & $0.96$ & $1.01$ \\
         Unconstrained MCFM (N=500) & $1.01$ & $0.97$ \\
         APPR (N=10000) & $0.94$ & $0.95$ \\
         APPR (N=500) & $0.93$ & $0.94$ \\        
         Jamming (N=2000) & $\gtrsim 1.00$ & $---$ \\
         Constrained MCFM (N=500) & $1.04$ & $1.23$ \\
         INEP (N=500) & $0.44$ & $0.56$ \\
\bottomrule
\end{tabular}
    \caption{Normalized differences $\delta_{4,3}$ and $\delta_{5,4}$ for each model.
    These values quantify how the widths of the 3-to-4 and 4-to-5 $k$-core transitions deviate from the Erdős–Rényi benchmark.}
    \label{tab:delta_kcores}
\end{table}

\begin{figure} [H]
    \centering
    \subfloat[]{
   \label{UCMCFM}
    \includegraphics[width=0.47\textwidth,height=6.5cm]{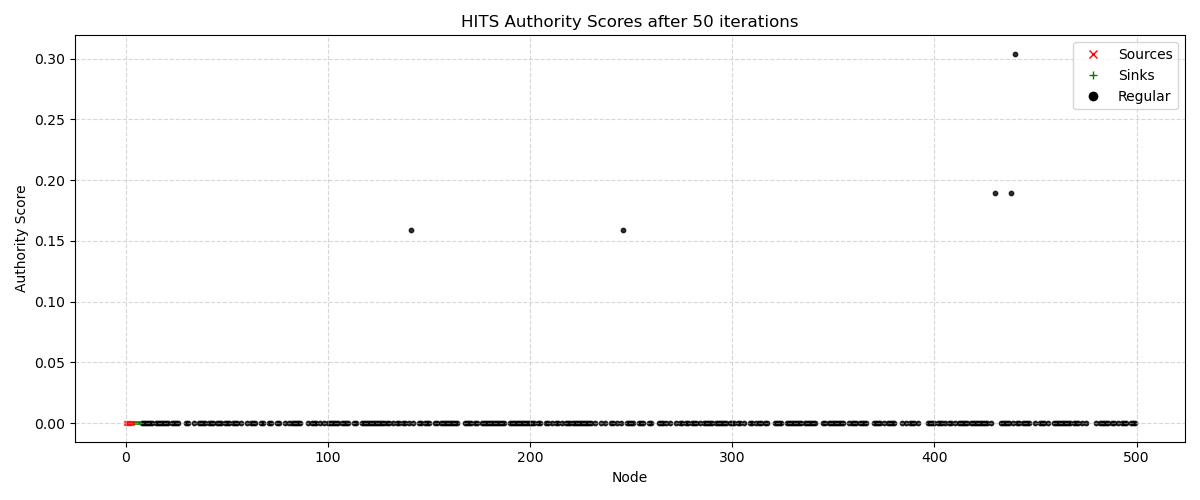}}\\
  \subfloat[]{
   \label{INEPS}
    \includegraphics[width=0.47\textwidth,height=6.5cm]{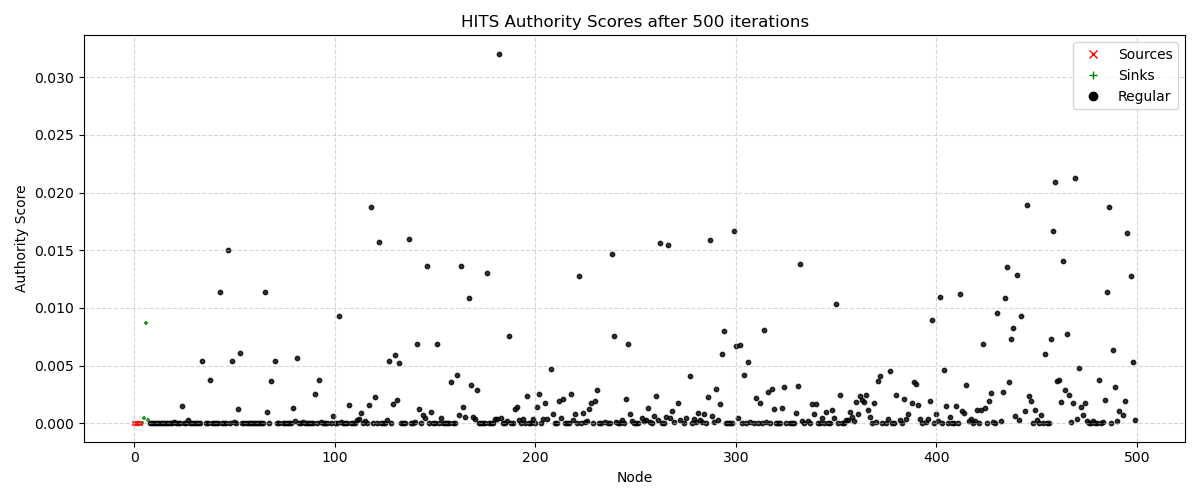}}\\
    \subfloat[]{
   \label{2core}
    \includegraphics[width=0.47\textwidth,height=6.5cm]{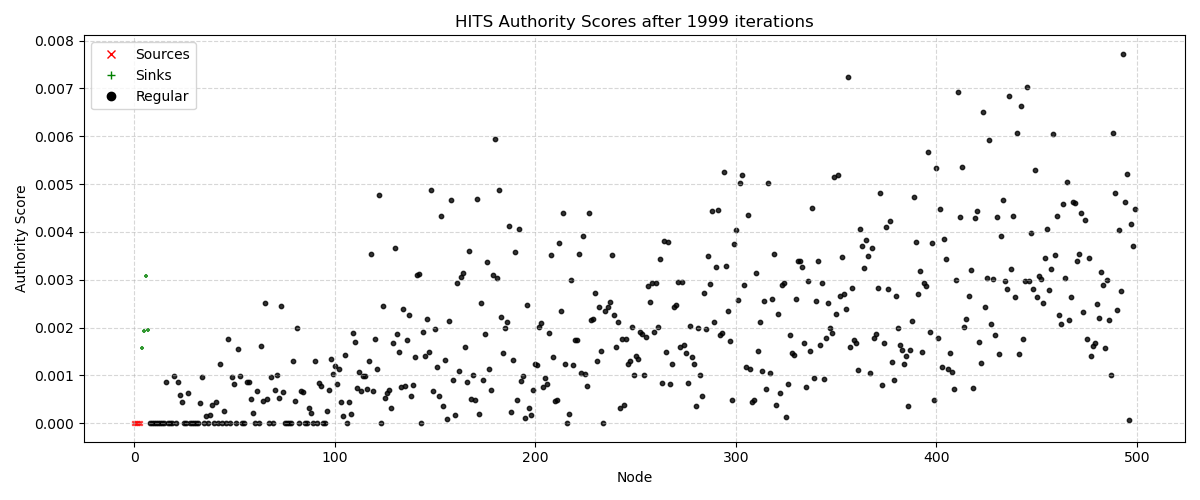}}\\
    \caption{Authorities scores at different stages of the evolution of the Unconstrained MCFM.}
    \label{fig:UMCFMauths}
\end{figure}

\begin{figure} [H]
    \centering
    \subfloat[]{
   \label{UCMCFM}
    \includegraphics[width=0.47\textwidth,height=6.5cm]{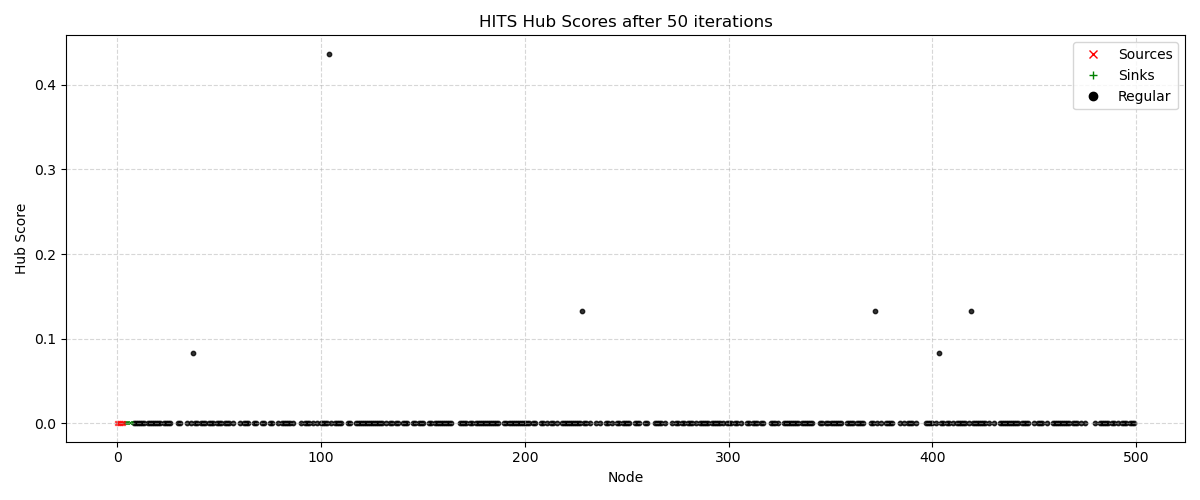}}\\
  \subfloat[]{
   \label{INEPS}
    \includegraphics[width=0.47\textwidth,height=6.5cm]{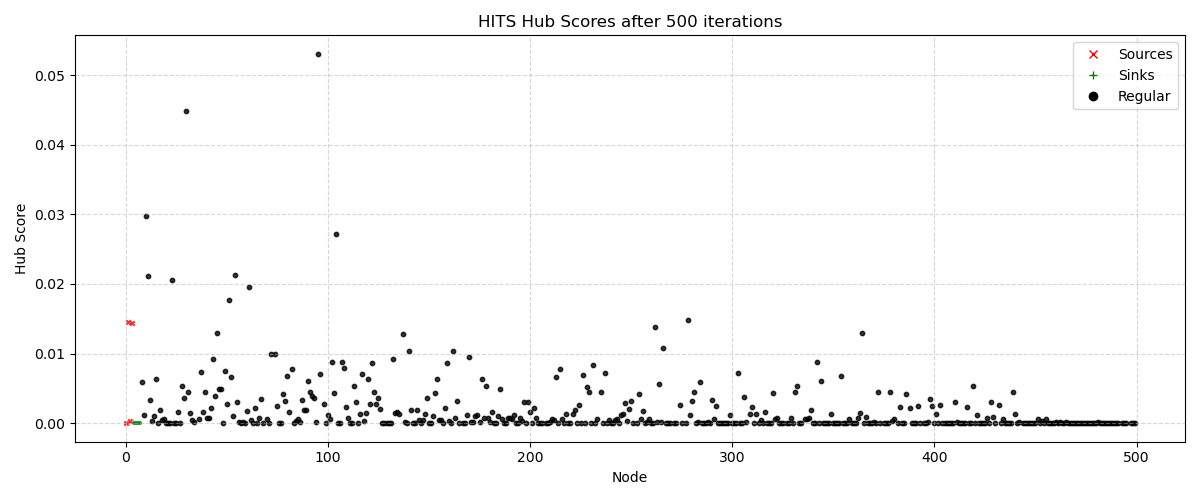}}\\
    \subfloat[]{
   \label{2core}
    \includegraphics[width=0.47\textwidth,height=6.5cm]{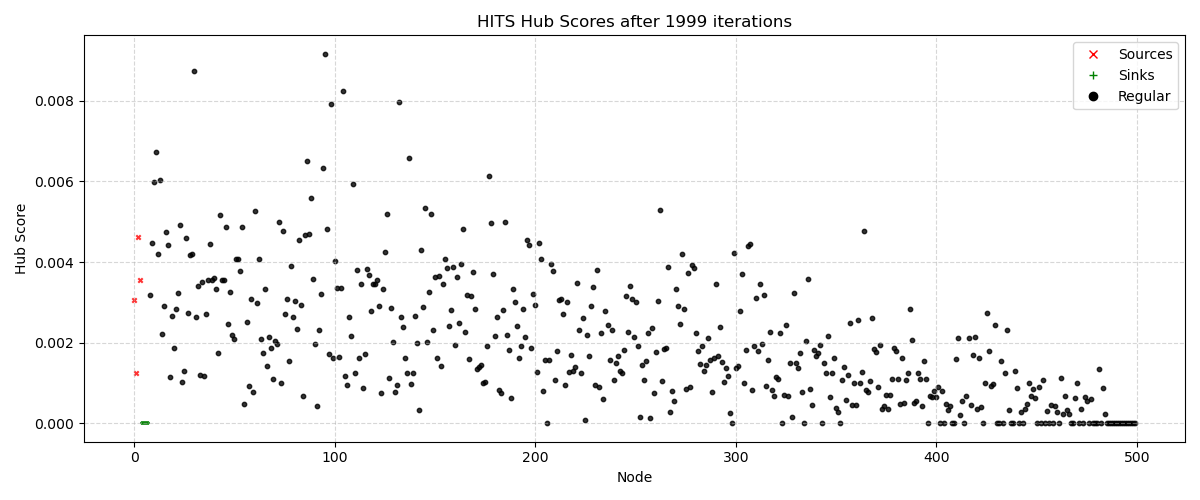}}\\
    \caption{Hubs scores at different stages of the evolution of the Unconstrained MCFM.}
    \label{fig:UMCFMhubs}
\end{figure}

\begin{figure} [H]
    \centering
    \subfloat[]{
   \label{CMCFM}
    \includegraphics[width=0.47\textwidth,height=6.5cm]{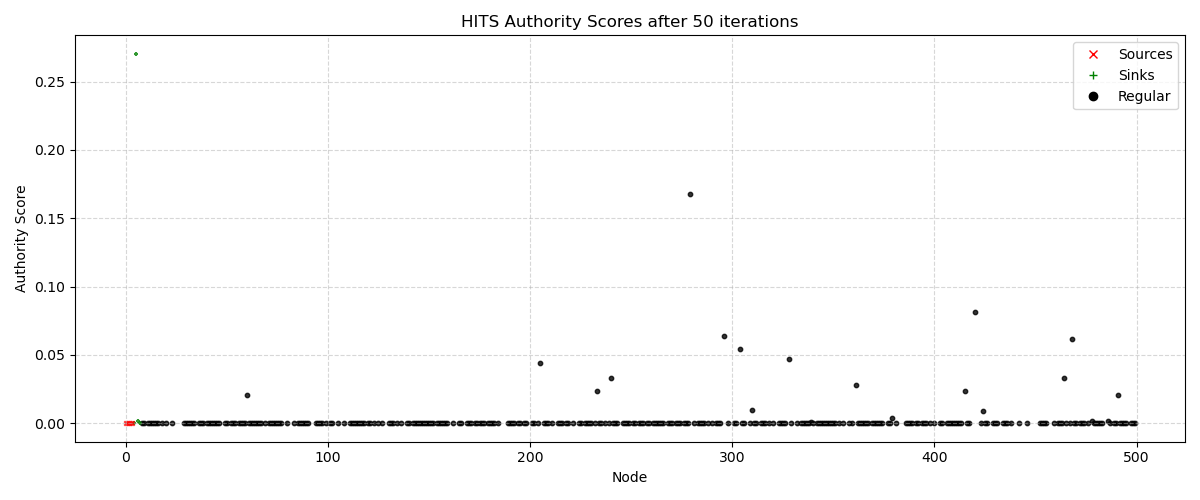}}\\
  \subfloat[]{
   \label{INEPS}
    \includegraphics[width=0.47\textwidth,height=6.5cm]{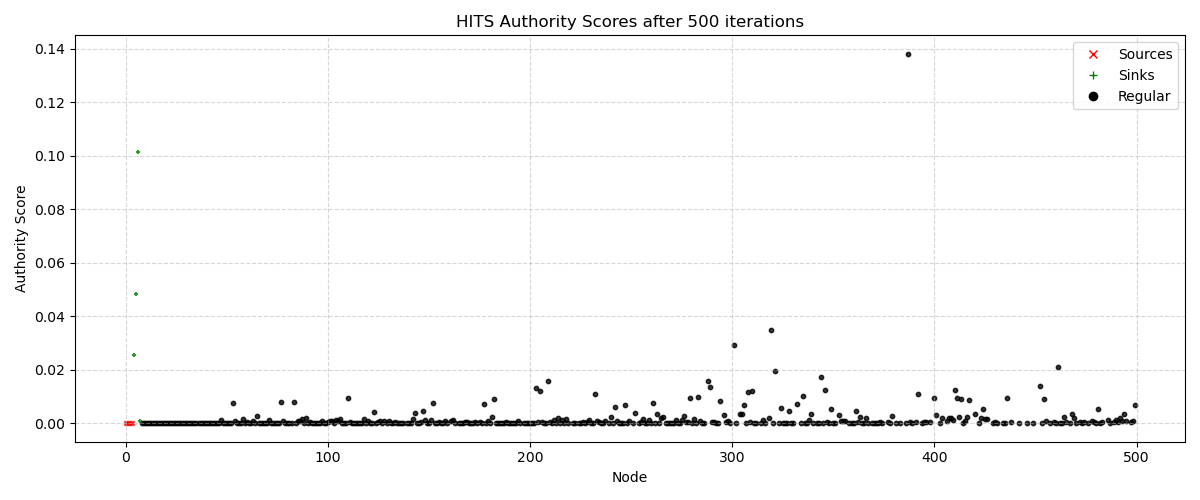}}\\
    \subfloat[]{
   \label{2core}
    \includegraphics[width=0.47\textwidth,height=6.5cm]{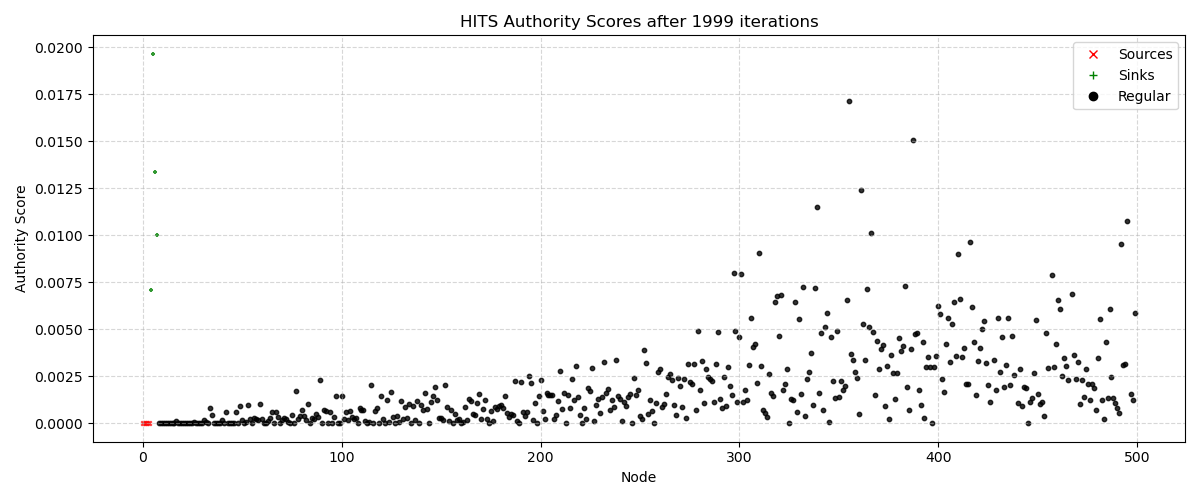}}\\
    \caption{Authorities scores at different stages of the evolution of the Constrained MCFM.}
    \label{fig:CMCFMauths}
\end{figure}

\begin{figure} [H]
    \centering
    \subfloat[]{
   \label{CMCFM}
    \includegraphics[width=0.47\textwidth,height=6.5cm]{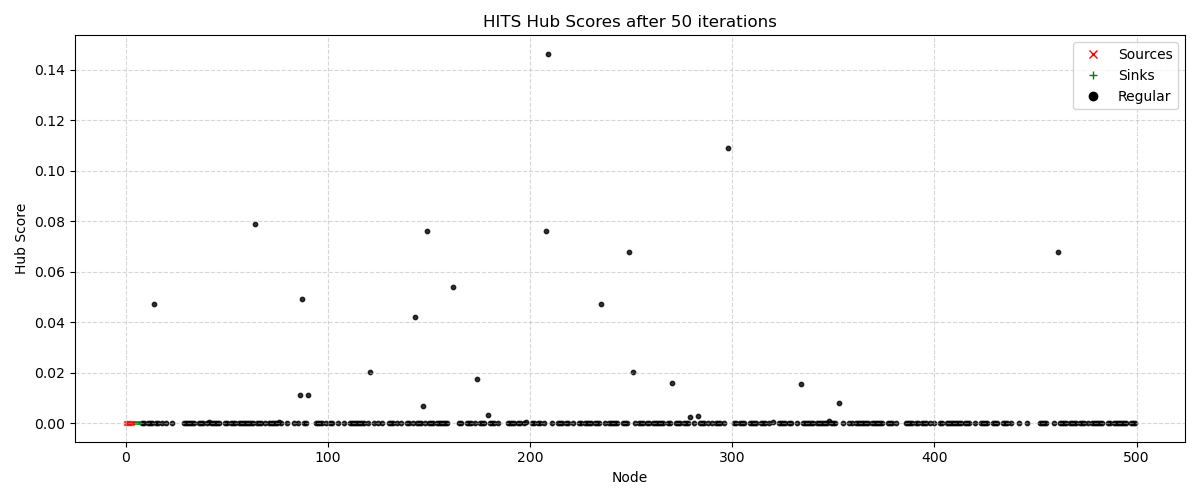}}\\
  \subfloat[]{
   \label{INEPS}
    \includegraphics[width=0.47\textwidth,height=6.5cm]{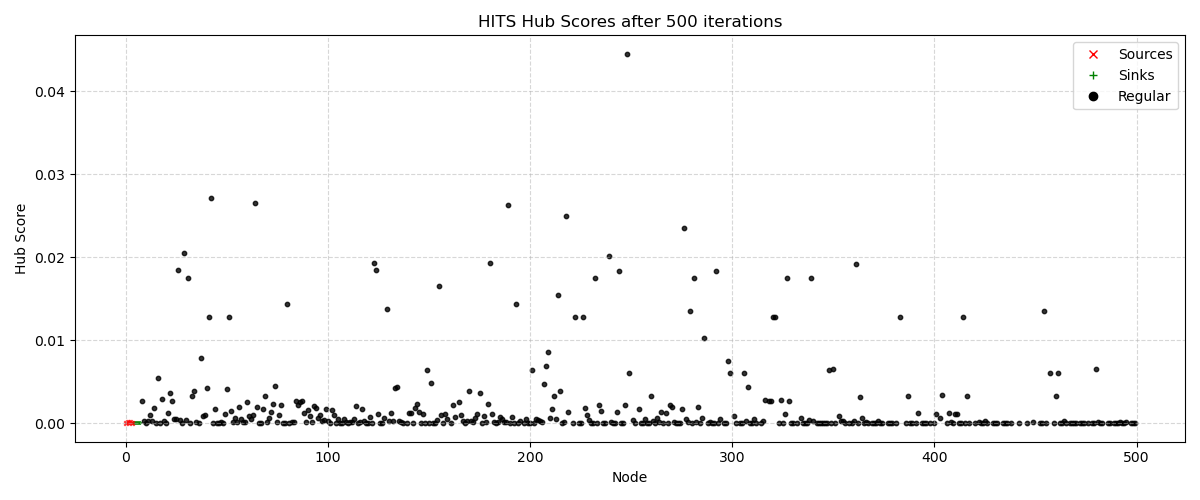}}\\
    \subfloat[]{
   \label{2core}
    \includegraphics[width=0.47\textwidth,height=6.5cm]{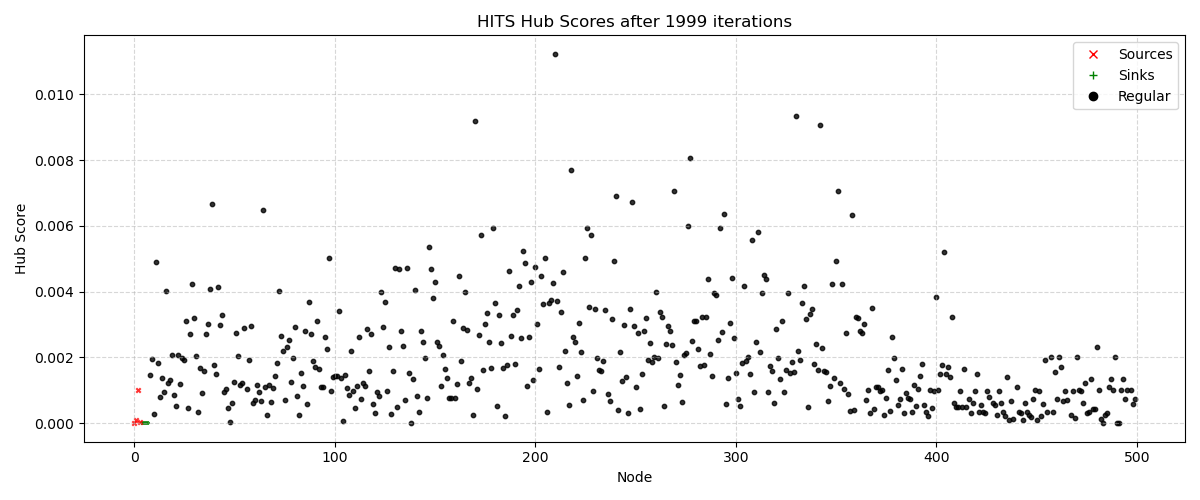}}\\
    \caption{Hubs scores at different stages of the evolution of the Constrained MCFM.}
    \label{fig:CMCFMahubs}
\end{figure}

While the above metrics capture the global pace of structural emergence, 
they do not directly address the presence of dominant nodes or pathways. 
To this end, we next examine hub formation and node centrality,
with particular attention to differences across models in terms of degree distribution and HITS scores
\cite{kleinberg1999}.
By definition, 
the ER model does not produce hubs; 
its degree distribution is narrow and homogeneous. 
In contrast, 
hubs emerge naturally in the PA model due to its reinforcement mechanism, 
though its undirected nature precludes a meaningful interpretation of authority versus hub roles. 
Similar limitations arise in the APPR, the Jamming and clustered models, 
where directionality is either absent or not structured in a way that supports HITS-based interpretation.
Moreover, 
because the remaining models incorporate stochastic elements, 
the HITS scores of individual nodes can vary significantly across simulations. 
As a result, 
averaging these scores is not informative. 
Instead, we present here representative examples from single runs of each model.
For the unconstrained MCFM,
snapshots of the change of the HITS scores
are shown in Figs.~\ref{fig:UMCFMauths} and \ref{fig:UMCFMhubs}.
As mentioned before,
SA nodes are $\{0, \cdots, 7\}$,
so they are distributed near the origin of the horizontal axis.
Sources are marked with $\times$ and sinks with $+$
(in
Table~\ref{tab:hubs_mcfm_ns} of the appendix \ref{app:hubs_authorities}
are reported the top 20 nodes by degree, 
hub score, 
and authority score by the end of the simulation). 
Hub and authority scores are often sparse or asymmetric in this case, 
reflecting the random nature of connections.
By contrast, 
the constrained MCFM 
(Figs.~\ref{fig:CMCFMauths} and \ref{fig:CMCFMahubs})
generates more structured connectivity, 
leading to distinct hub and authority patterns 
(check also Table~\ref{tab:hubs_mcfm_s} in appendix \ref{app:hubs_authorities}).

\begin{figure} [H]
    \centering
    \subfloat[]{
   \label{INEP}
    \includegraphics[width=0.47\textwidth,height=6.5cm]{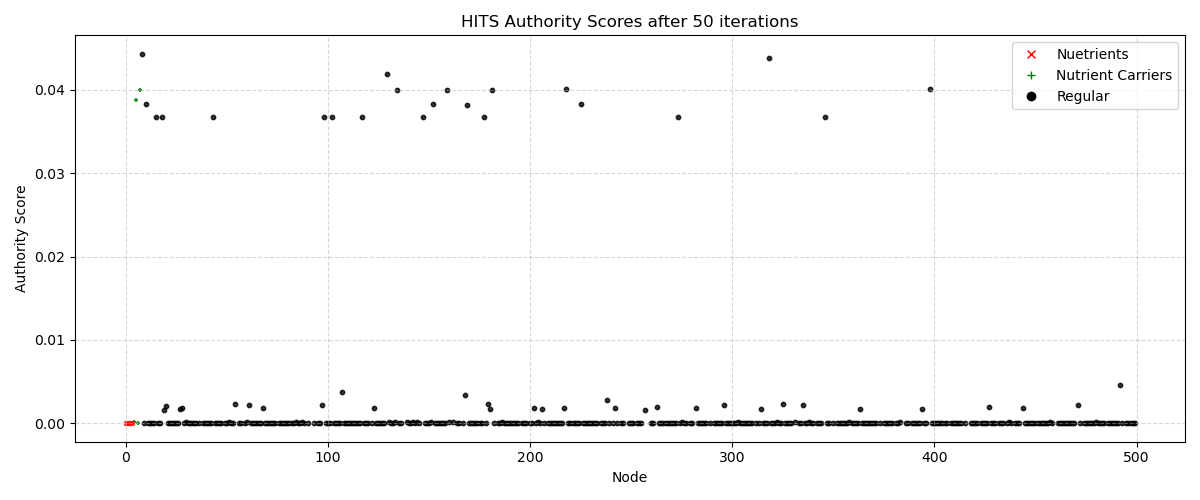}}\\
  \subfloat[]{
   \label{INEPS}
    \includegraphics[width=0.47\textwidth,height=6.5cm]{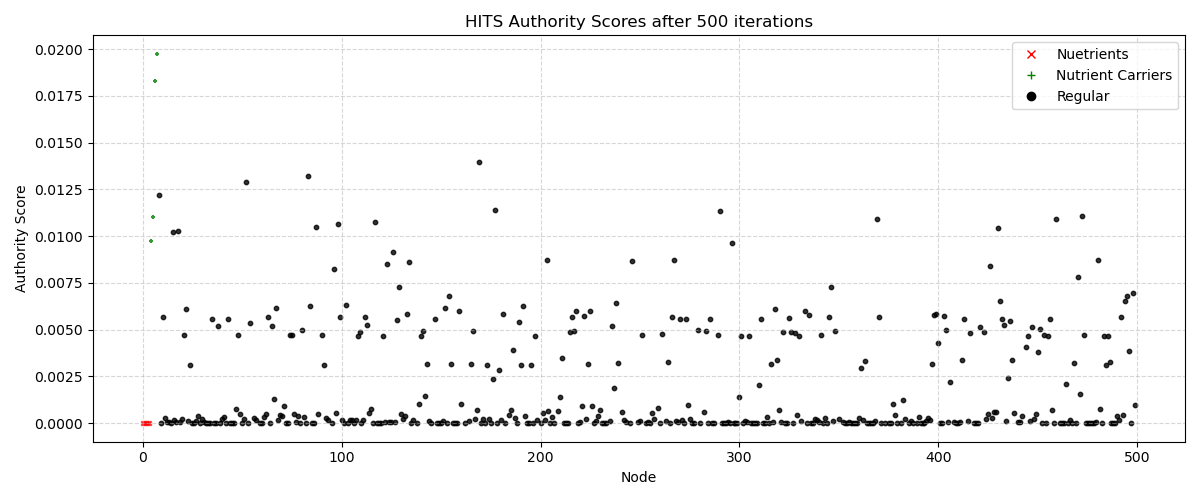}}\\
    \subfloat[]{
   \label{2core}
    \includegraphics[width=0.47\textwidth,height=6.5cm]{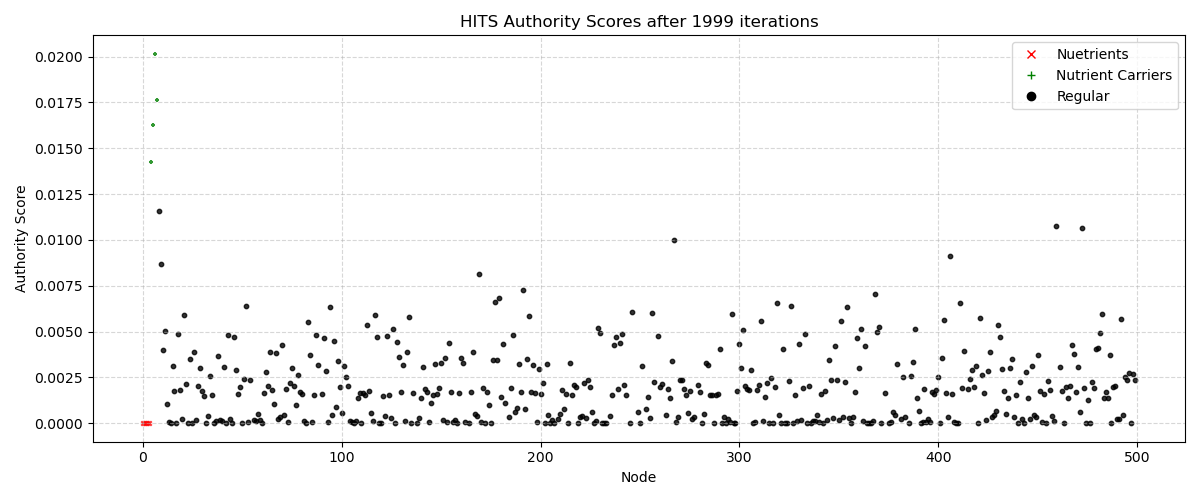}}\\
    \caption{Authorities scores at different stages of the evolution of the for INEP.}
    \label{fig:INEPauths}
\end{figure}

\begin{figure} [H]
    \centering
    \subfloat[]{
   \label{CMCFM}
    \includegraphics[width=0.47\textwidth,height=6.5cm]{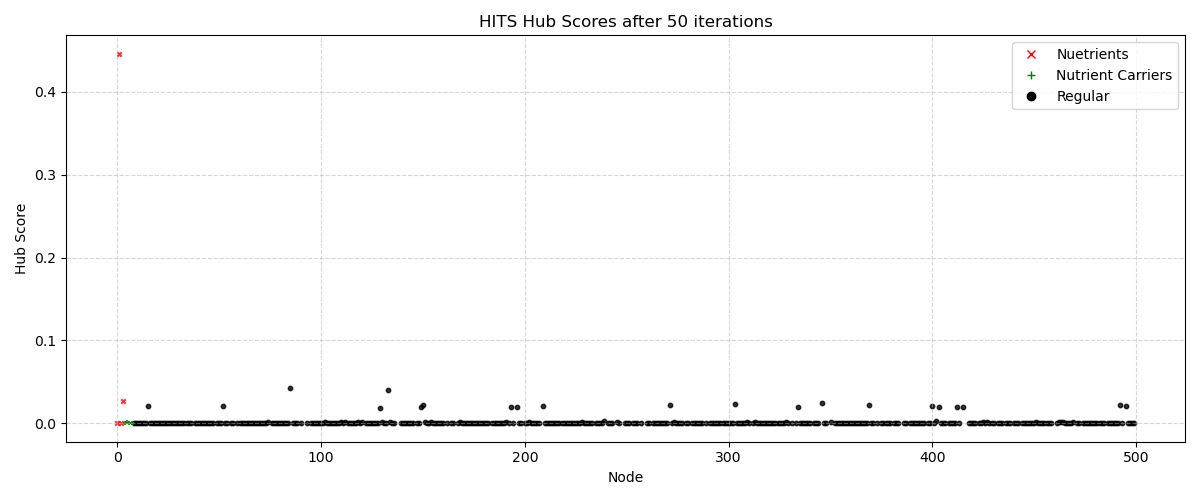}}\\
  \subfloat[]{
   \label{INEPS}
    \includegraphics[width=0.47\textwidth,height=6.5cm]{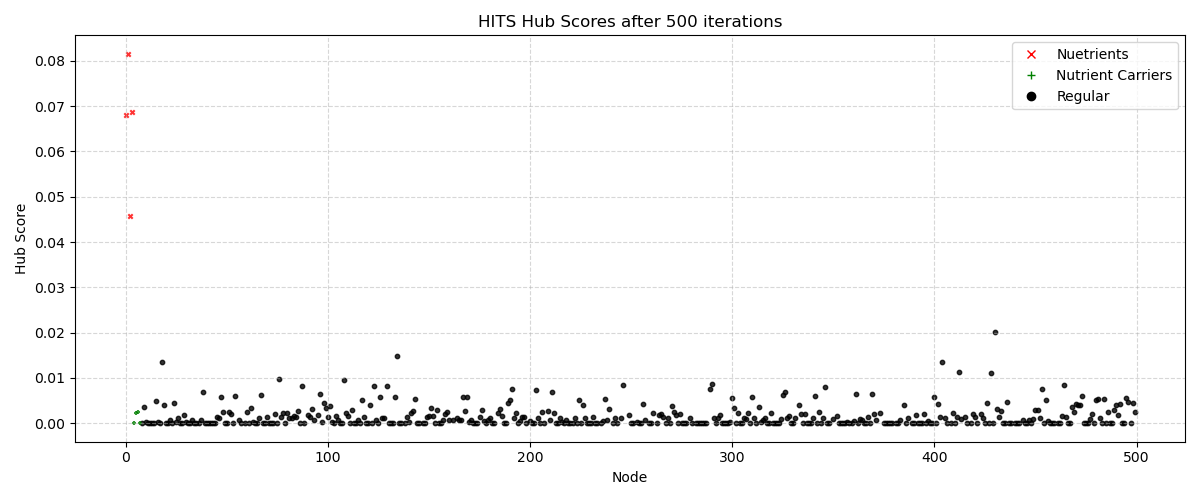}}\\
    \subfloat[]{
   \label{2core}
    \includegraphics[width=0.47\textwidth,height=6.5cm]{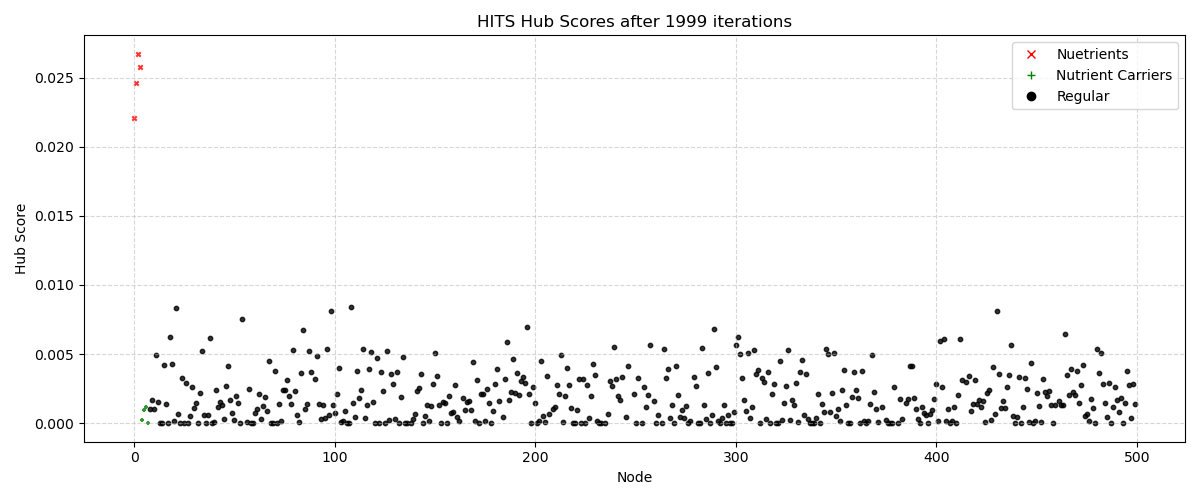}}\\
    \caption{Hubs scores at different stages of the evolution of the Constrained INEP.}
    \label{fig:INEPhubs}
\end{figure}

The INEP stands out due to its explicit duplication-growth mechanism. 
In Figs.~\ref{fig:INEPauths} and \ref{fig:INEPhubs},
the eight SA 
(nutrients, $\times$; nutrients carriers, $+$)
are distributed near the origin of the horizontal axis too.
Table~\ref{tab:hubs_cem} in appendix \ref{app:hubs_authorities}
shows that,
at the end of the simulation,
a few nodes possess disproportionately high degrees and centrality scores, 
which is consistent with the model’s growth and duplication rules.

\newpage



\section{Discussion}
\label{sec:discuss}

The results reported in the previous section 
shed light on the interplay between topology and functionality in network-based complex systems.
A key insight guiding our comparative analysis is that, 
in finite-size networks, 
all link-adding processes eventually converge to a complete graph. 
Since all complete graphs are structurally equivalent, 
the long-term global order resulting from different processes become indistinguishable. 
Therefore, 
meaningful differences between finite-size models can only be observed during their transient regimes,
that is, 
before convergence to the complete graph erases structural diversity. 

The first thing to notice in Table~\ref{tab:av_degree} 
is that,
with the predictable exception of the APPR,
the formation of the 2-core seems to occur at $\langle k \rangle$ close to $1$ across all models,
even in processes with strong selection pressures, 
such as the constrained MCFM
or with active distinct agents
as, for instance, nodes with highest connectivity in the PA model.
Moreover,
when adding links completely randomly, 
the rising of the 2-core is known to happens gradually
\cite{pittel1996}, 
and in the top panels of figures \ref{fig:uMCFMkcores} and \ref{fig:cMCFMkcores}
it can be observed that this seems to be true for $k=2$,
even for models with SA
(sources, sinks)
and GSR
(higher sink absorption).

A fundamental constraint shared by most link-adding processes on finite-size networks is the combinatorial bottleneck imposed by sparse connectivity. 
For example, consider a graph with $6M$ vertices and $2M$ edges, 
arranged as $2M$ disconnected pairs. 
Such a graph has evolved through $2M$ iterations of edge additions, 
yet remains structurally incapable of supporting a 2-core: 
every node has degree at most 1, 
and the average degree is $2/3$. 
In this regime, 
most of the functionally guided or selection-driven strategies,
provided they still include a degree of randomness,
will typically not produce nontrivial core structures, 
simply because no local neighborhood achieves the minimal density required. 
This limitation generalizes to broader classes of sparse networks and highlights a crucial insight: 
in early stages of evolution, topology dominates strategy. 
Generically,
functional pressures, 
selection rules, 
or agent-specific roles 
will only accelerate core formation once a minimal substrate of structural complexity is present. 
Thus, 
given some randomness in the link-addition policy, 
sufficiently sparse networks will usually evolve similarly to large-$N$ ER processes in their early phase, 
until the accumulation of links crosses a threshold beyond which functional organization can exert meaningful influence.

In this context, 
the atypical behavior of the APPR provides particularly compelling evidence. 
Among all models examined, 
APPR is the only one in which the rise of the 2-core is significantly delayed, 
occurring at $\langle k\rangle \approx 1.60$,
well above the expected threshold for random linking. 
This shift supports the idea that the GSR in APPR
(designed to suppress the formation of large components)
acts as a hindrance to reaching the minimal structural substrate necessary for 2-core emergence. 
Unlike other models where functional biases or local preferences operate alongside randomness, 
the deterministic side of APPR’s selection mechanism actively steers the system away from topological configurations that could seed a 2-core. 
In this sense, 
APPR exemplifies how a sufficiently strong GSR can prevent the spontaneous emergence of non-trivial structure, 
underscoring the importance of early topological conditions in enabling the action of evolutionary pressures.
However, 
this apparent suppression comes with a trade-off: 
the threshold for 2-core emergence in APPR is accompanied by a markedly higher standard deviation than those of higher $k$-cores, 
and higher than what is typically observed in other models. 
This variability indicates that, 
although the average onset is delayed, 
the actual transition point fluctuates significantly across realizations, 
reflecting a high sensitivity to microscopic fluctuations in early configurations. 
Moreover, 
this variability aligns with the explosive nature of the GCC in Achlioptas processes,
where connectivity is actively discouraged until a critical point is reached, 
beyond which a large connected component emerges abruptly. 
The sudden appearance of the 2-core thus mirrors the known percolation dynamics of APPR
\cite{achlioptas2009}, 
suggesting that what appears as a delayed structural onset is in fact a compressed, 
sharply synchronized transition,
a hallmark of explosive phenomena. 
In this view, 
the high standard deviation is not a weakness of the mechanism 
but a signal of the strongly nonlinear regime governing the system’s response near the threshold.

Now,
from Table~\ref{tab:av_degree}
it can be observed that
in some models like APPR (GSRnSA),
Jamming (GSRnSA),
and unconstrained MCFM (nGSRSA), 
after the rise of the 2-core,
the network keeps evolving analogously to an ER random graph (nGSRnSA),
clearly indicating that, 
even with the condition of minimal complexity satisfied,
the presence of GSR or SA 
is not a sufficient condition for the $k$-cores to arise faster 
than in a purely random process.
For instance, 
in the unconstrained MCFM,
while sources and sinks are structurally defined, 
they remain functionally neutralized due to the absence of a selection mechanism. 
As it can be observed in Figs.~\ref{fig:UMCFMauths}, \ref{fig:UMCFMhubs} and Table \ref{tab:hubs_mcfm_ns},
for this case the degree distribution remains narrow, 
hubs are always absent, 
and both HITS hub and authority scores keep diffuse and unremarkable. 
The network’s evolution lacks directionality: 
new links are added randomly, 
and structure-function alignment does not emerge.

\begin{figure}[h]
     \centering
     \includegraphics[scale=0.5]{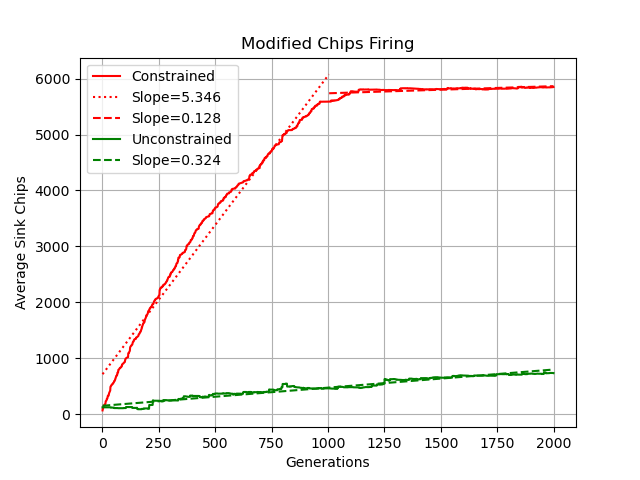}
     \caption{Examples of Chips-Firing dynamics.
     The vertical axis stands for the average amount of chips absorbed by the sinks at each generation of the process.
     The top (red) curve corresponds to a constrained model.
     For this case two linear fits are shown, before and after reaching a plateau.
     The bottom (green) curve correspond to an unconstrained model.
     The slope of the corresponding linear fit is also shown.}
     \label{fig:sinkchips}   
\end{figure}

That is closely related to what can be observed,
by comparing Tables~\ref{tab:av_degree} and \ref{tab:delta_kcores},
for models PA (nGSRSA) and constrained MCFM (GSRSA).
Though the 3-core undoubtedly arises earlier than in the ER model,
higher $k$-cores appear at ratios comparable to those in purely random process.
Focusing on the data for the constrained MCFM 
in Figs.~\ref{fig:CMCFMauths} and \ref{fig:CMCFMahubs}
and Table \ref{tab:hubs_mcfm_s}, 
we observe a sharper deviation from random growth. 
Sink nodes get to dominate authority rankings, 
and a more differentiated topology emerges, 
including intermediate hubs that route flows efficiently. 
These features suggest that 
in the earlier stages
selection promotes a functionally specialized architecture. 
The example for the sink chip absorption dynamics shown in Figure~\ref{fig:sinkchips}, 
illustrate this: 
early in the simulation, chips are absorbed at a much higher rate (slope $= 5.35$) 
than for the unconstrained case ($0.32$), 
reflecting rapid functional optimization. 
However, 
this growth eventually \textbf{plateaus} 
(new slope $= 0.13$), 
indicating that while early selection drives strong improvements, 
the system hits a ceiling. 
Notably, 
this transition coincides with the observation that higher $k$-core emergence slows down 
and begins to mirror the growth rate seen in ER graphs. 
This saturation dynamic is crucial. 
It suggests that selection alone,
if focused on a single scalar functionality like sink absorption,
may be insufficient to maintain structural innovation indefinitely. 
Once a minimal efficient configuration is reached, 
the network ceases to evolve structurally despite continued selection. 
This is a core limitation also visible in the PA model, 
where hubs dominate but clustering remains low, 
and deeper $k$-cores fail to emerge faster than in a purely random process.

The saturation just discussed can be mitigated by increasing the selective pressure imposed by the GSR. 
In Achlioptas-like processes, 
such as those in rows 9 to 11 of Table~\ref{tab:av_degree}, 
increasing the probability $q$ of selecting edges that connect vertices already linked to a common neighbor 
(thus promoting clustering) 
from 0.01 to 0.5 significantly advances the appearance of the 3-core, 
shifting it from near the ER threshold to much earlier stages of the evolution. 
Although direct data for higher $k$-cores is not available for these models, 
the earlier emergence of the 3-core suggests that sufficiently high values of $q$ 
may also accelerate the formation of higher-order $k$-cores, 
which typically build upon lower-order ones.
In this context, 
the INEP stands out as a notable exception. 
Unlike Achlioptas-like processes that rely on carefully calibrated GSR to accelerate core formation, 
INEP achieves comparable structural outcomes
without requiring fine-tuning. 
Here, 
the selection rule is aligned with growth functionality,
favoring faster-dividing cells,
and the intracellular network must support complex biochemical transformations through locally redundant reaction subgraphs. 
As can be observed in Fig.~\ref{fig:cellkcores},
this dual pressure fosters the formation of deep, 
nested $k$-cores
formed faster than in purely random process
(with the possible exception of $k=2$).
The values for $k\geq 3$ can be seen in Table~\ref{tab:av_degree} 
and the deviation from ER in Table~\ref{tab:delta_kcores}.
Moreover,
we also found for this process that
$\left<k\right>_6=5.22\pm 0.27$,
and $\left<k\right>_7=6.40\pm 0.30$,
with the corresponding normalized differences given by
$\delta_{6,5}=0.65$ and $\delta_{7,6}=0.78$.
So that,
in the INEP higher $k$-cores keep rising swiftly,
even considering the upper bound for $\left<k\right>_7$,
it is still lower than $\left<k\right>_5$ for ER.
On the other hand,
as seen in the top-bottom sequences in Figs.~\ref{fig:INEPauths} and \ref{fig:INEPhubs},
and Tab.~\ref{tab:hubs_cem},
the topology-functionality alignment gives rise to extreme centralization in both degree and HITS centrality.

An immediate inference from the discussion so far 
is that the ER process serves as a natural null model: 
it represents the most unbiased link-addition process. 
Since even relatively small ER networks exhibit the same self-organization dynamics as their large-$N$ counterparts
(see for instance the threshold values for $N=500$ in the third row of Table~\ref{tab:av_degree}), 
this justifies using the ER trajectory as a baseline against which to assess other finite-size models. 
Specifically, 
it allows us to examine 
(i) whether a given model deviates from the ER trajectory, 
(ii) how early and strongly such a deviation occurs, 
and (iii) how long the model resists convergence toward ER-like topology. 
These comparisons form the backbone of our analysis of structural and functional differentiation in network-based random processes.
Useful clues for detecting this deviation are the emergence of the k-cores
(values in the first row of Table~\ref{tab:av_degree}). 
Models that self-organize beyond any of these points faster or slower than the ER process can be said to exhibit strategy-driven organization. 
Conversely, those that converge to the same topological milestones as ER graphs
(regardless of their GSR or SA)
are effectively indistinguishable from random processes during the corresponding phase of their evolution. 

Together with the rise of specialized hubs,
this framing offers a unified and quantitative approach to evaluate the impact of selection and agent dynamics 
across different classes of network-based random processes.
Let us compare, 
under this perspective,
the results for the weakly and strongly connected susceptibilities
for the two versions of MCFM
and the INEP,
all of them with topology given by a directed graph
and functionality intended to redistribute across the system
resources taken from the environment.
The corresponding plots are displayed in the middle and bottom panels of 
Figs.~\ref{fig:uMCFMkcores}, \ref{fig:cMCFMkcores}, and \ref{fig:cellkcores}. 
A peak in $S(\left<k\right>)$ typically signals the onset of large-scale reorganization, 
akin to a phase transition in physical systems, 
and marks the critical regime where core-like connectivity begins to emerge.
For all models, 
$S_w(\left<k\right>)$ exhibits a clear peak near the point where each $k$-core emerges, 
confirming that the system's capacity to support large undirected substructures
increases as links are added. 
Again,
this behavior mirrors the classical picture observed in Erdős–Rényi networks 
\cite{newman2003}
and serves as a minimal indicator of global structural readiness.
However, 
a striking difference appears when directionality is taken into account. 
For both the unconstrained and constrained MCFM 
(Figs.~\ref{fig:uMCFMkcores} and \ref{fig:cMCFMkcores}), 
even if their topologies are given by directed graphs,
the strongly-connected susceptibility remains flat, 
at $S_s(\left<k\right>) = 1$ 
across the entire evolutionary trajectory. 
This plateau indicates that the non-core portion of the graph remains composed solely of trivial strongly-connected components
(mostly isolated vertices or acyclic structures)
with no formation of loops or feedback-rich motifs. 
Despite the presence of selection mechanisms in the constrained version, 
there is no evidence of global organization into strongly-connected subgraphs beyond the emergent core. 
This suggests that although selection may enhance undirected cohesion, 
it fails to generate meaningful cyclic or integrative feedback at the network-wide scale.
In sharp contrast, 
the INEP 
(Fig.~\ref{fig:cellkcores}) 
displays a markedly different behavior: 
the strongly-connected susceptibility $S_s(\left<k\right>)$ shows six distinct peaks, 
each aligned with the emergence of a strongly-connected $k$-core. 
This reflects a capacity to not only increase undirected connectivity, 
but also to systematically build cyclic, 
functionally integrated structures in a hierarchical fashion. 
In the INEP,
the synergy between GSR and SA creates a self-reinforcing feedback loop: 
topologies that promote functionality are favored, 
and functional performance depends on increasingly complex topological features.
The behavior of the centrality rankings in the INEP shown in figures \ref{fig:INEPauths} and \ref{fig:INEPhubs}
further validate this interplay
(see also Table \ref{tab:hubs_cem}). 
Nutrient nodes 
(nodes 0–3) 
dominate degree and hub scores due to their pervasive role in initiating reactions. 
Nutrient carriers 
(nodes 4–7), 
although not initiating reactions, 
accumulate high authority scores due to their centrality in sustaining nutrient gradients. 
This clear structural-functional correspondence
(hubs as sources, authorities as regulators)
highlights how \textbf{topological differentiation} emerges directly from \textbf{functional necessity}.

A particularly illuminating contrast arises when we compare the behavior of sources in the MCFM 
with that of nutrients in the INEP. 
At first glance, 
both serve as inputs driving the system: 
sources inject chips in MCFM, 
while nutrients are imported from the environment in INEP. 
However, 
their functional roles within the respective dynamics are fundamentally different. 
In the MCFM, 
sources act unconditionally, 
firing chips at each iteration regardless of the network state. 
Once connected, 
they continue to push chips into the system, 
but their influence is largely constrained by the network’s capacity to channel chips toward sinks. 
Additional connectivity of sources does not necessarily improve global performance; 
in fact, 
it may create bottlenecks if not matched by complementary paths. 
In stark contrast, 
INEP’s nutrient nodes are embedded within a network of feedback dependencies: 
nutrients can only be replenished if consumed, 
and consumption requires catalytic activity from other species that themselves depend on nutrients for growth. 
This interdependence creates loops in which input availability is tightly coupled to internal organization. 
Furthermore, 
the emergence of power-law concentration distributions in INEP 
\cite{furusawa2003,radillo2023}
reflects a system in which input nodes dominate quantitatively, 
yet remain functionally entangled 
at some degree with most of the species. 
These differences underscore the importance of feedback loops and interdependency in enabling self-organization, 
as opposed to the more rigid, 
feedforward structure of source-driven models like the MCFM.

Let us finish this section by paying attention to the fact that,
as shown in the bottom curve of Fig.~\ref{fig:sinkchips},
in the unconstrained MCFM
the average number of chips absorbed by sinks also increases linearly, 
thought with a modest slope. 
This gradual increase confirms that even in the absence of functional guidance, 
connectivity growth alone suffices to improve transport, 
albeit inefficiently. 
The chips reach the sinks as a byproduct of expanded random connectivity,
supporting the idea that even random networks accumulate functionality over time due to structural accretion. 
Interestingly, 
the spontaneous emergence of functional structure in purely stochastic network evolution bears a conceptual resemblance to the \textit{placebo effect} in medicine. 
In both cases, 
observable improvements arise not through targeted intervention or optimized design, 
but as a side effect of internal dynamics. 
Just as a placebo elicits measurable physiological responses through the belief in treatment 
or the natural predisposition of the patient's immune system, 
a randomly evolving network can exhibit 
the formation of weak $k$-cores
and increasing transport efficiency 
purely due to the statistical consequences of accumulating edges. 
This passive gain in functionality, 
absent any selection mechanism, 
may misleadingly resemble purposeful organization. 

\section{Conclusions}
\label{sec:concl}

It has been known
that during the evolution of graph-based random processes
(i.e., network-based random processes with trivial functionality), 
highly interconnected structures and core-periphery hierarchies can arise purely by chance. 
In this work,
by studying a modified version of the chips-firing model with special agents 
but without a selection mechanism, 
we observed that the emergence of $k$-cores and the formation of hubs 
(whether measured by degree or HITS authority scores) 
closely follows the behavior expected in an Erdős–Rényi random graph model.
Nevertheless,
in this new scenario, 
the gradual rise in sink absorption capacity reflects the random growth of interconnectivity, 
confirming that increased complexity alone can produce modest improvements in functionality 
even in the absence of goal-directed evolution.
In other words, in stochastic network evolution models, 
even when the rules are entirely random, 
the progressive increase in connectivity leads to the spontaneous emergence 
of non-trivial structural features 
(such as densely connected subgraphs and weak core hierarchies),
alongside a gradual improvement in functional performance.
The strong analogy with the \textit{placebo effect} in medicine 
helps clarify that 
the mere appearance of order or efficiency in a complex system does not necessarily imply the presence of underlying goal-directed processes,
it may simply reflect the inevitable byproducts of connectivity growth in finite systems.

Though 
purely random process are unguided and inefficient, 
lacking any optimization toward specific functional outcomes,
the above conclusion is valid for any link-adding process on a finite-size network.
In such cases,
the asymptotic topological configuration is a complete graph,
implying that,
at a given moment of the network evolution,
its topology will change along the lines of an Erdős-Rényi dynamics.
We found that
to observe significant deviations from purely random-network trajectories  
(either globally or during particular phases of evolution) 
in random processes with some degree of determinism
over sufficiently large but finite networks, 
two key conditions must be satisfied simultaneously:
\begin{enumerate}
    \item \textbf{Critical complexity:} 
    To allow functionally meaningful differentiation,
    the network must either contain sufficient local redundancy or clustering,
    or exhibit a suppression of such structures beyond what would arise by chance.
    Without structural motifs such as cycles, cliques, or tightly knit neighborhoods,
    local and global selection pressures have little substrate to act upon,
    unless they act specifically to hinder the emergence of that complexity.
    \item \textbf{Topology–functionality alignment:}
    There must be a dynamic feedback between the evolving network topology
    and the system’s functional goals,
    such that local or global selection mechanisms can preferentially reinforce
    or suppress
    structural configurations
    that enhance or inhibit functionality 
    (e.g., efficient transport, robust growth, or adaptive regulation).
\end{enumerate}
If either of these conditions is absent, 
most evolutionary processes remains statistically indistinguishable 
from that of a random graph, 
regardless of whether global selection rules or special agents are present in name.
Therefore,
as criteria for the fulfillment of these conditions
can serve
the comparison
with the known critical values for the rise of $k$-cores in the large-$N$ Erdős–Rényi model
(both, of magnitudes and of their normalized differences),
as well as the emergence of specialized hubs.

Our findings also have broader implications for both the interpretation of natural complex systems and the design of artificial ones.
If a real-world system with known selection rules and special agents exhibits behavior
(as described by $k$-core emergence, weakly and strongly connected susceptibilities or authorities scores) 
that are similar to those of an Erdős–Rényi model, 
this suggests the futility of the special agents,
the lack of effectiveness of the selection rules, 
or both. 
In contrast, 
if a real-world system without explicitly known selection rules or special agents exhibits significant deviations from random graph behavior, 
this provides indirect evidence that,
though hidden,
such mechanisms may be present and functionally aligned. 
Thus, Erdős–Rényi-based benchmarks can serve as powerful diagnostic tools: 
they offer a null model against which to infer the presence, 
absence, 
or effectiveness of functional structuring mechanisms. 
This perspective may also aid in the design of engineered complex systems, 
by providing a baseline for evaluating whether topological evolution is truly being shaped by functional imperatives.

\section*{Acknowledgements}

The work of A. Puente was supported by CONAHCyT grant 1305175. 
The work conducted by D. Radillo was carried out with the support of the Swiss National Science Foundation (SNSF) grant 202775.
\newpage

\appendix
\section{Hub and Authority Score Tables}
\label{app:hubs_authorities}

For completeness, 
we provide here the numerical values of the hub and authority scores computed for selected networks at the end of our simulations. 
These metrics, derived from the HITS (Hyperlink-Induced Topic Search) algorithm \cite{kleinberg1999}, 
quantify the extent to which nodes function as centralized sources (hubs) or sinks (authorities) in the directed network representations used 
in three of our models.

\begin{table}[ht]
\centering
\caption{Top 20 nodes by degree, hub score, and authority score in the unconstrained MCFM (after 2000 edges, $N=500$).}
\begin{tabular}{|c|c|c|c|}
\hline
\textbf{Node} & \textbf{Degree} & \textbf{Hub Score} & \textbf{Authority Score} \\
\hline
256 & 19 & 0.000 & 0.006 \\
27 & 18 & 0.008 & 0.000 \\
394 & 18 & 0.000 & 0.006 \\
468 & 17 & 0.000 & 0.009 \\
380 & 17 & 0.000 & 0.008 \\
454 & 17 & 0.000 & 0.000 \\
497 & 17 & 0.006 & 0.006 \\
145 & 17 & 0.000 & 0.000 \\
377 & 17 & 0.000 & 0.006 \\
2   & 17 & 0.010 & 0.000 \\
190 & 16 & 0.000 & 0.000 \\
488 & 16 & 0.000 & 0.000 \\
170 & 16 & 0.006 & 0.000 \\
139 & 15 & 0.006 & 0.000 \\
129 & 15 & 0.007 & 0.000 \\
10  & 15 & 0.006 & 0.000 \\
368 & 15 & 0.000 & 0.006 \\
351 & 15 & 0.000 & 0.000 \\
35  & 15 & 0.000 & 0.000 \\
416 & 15 & 0.000 & 0.000 \\
\hline
\end{tabular}
\label{tab:hubs_mcfm_ns}
\end{table}

\begin{table}[ht]
\centering
\caption{Top 20 nodes by degree, HITS hub scores, and HITS authority scores in the constrained MCFM. Nodes 0--3 are sources, and nodes 4--7 are sinks.}
\begin{tabular}{|c|c||c|c||c|c|}
\hline
\textbf{Node} & \textbf{Degree} & \textbf{Node} & \textbf{Hub Score} & \textbf{Node} & \textbf{Authority Score} \\
\hline
6   & 29 & 315 & 0.010 & 7   & 0.019 \\
391 & 29 & 356 & 0.009 & 6   & 0.016 \\
7   & 28 & 398 & 0.009 & 391 & 0.015 \\
362 & 27 & 289 & 0.008 & 443 & 0.013 \\
354 & 24 & 199 & 0.008 & 468 & 0.012 \\
5   & 24 & 391 & 0.007 & 362 & 0.012 \\
408 & 23 & 329 & 0.007 & 354 & 0.011 \\
498 & 22 & 219 & 0.006 & 363 & 0.011 \\
470 & 21 & 337 & 0.006 & 394 & 0.010 \\
363 & 20 & 328 & 0.006 & 5   & 0.010 \\
392 & 19 & 436 & 0.006 & 408 & 0.010 \\
324 & 19 & 158 & 0.006 & 470 & 0.010 \\
219 & 19 & 62  & 0.006 & 463 & 0.010 \\
289 & 19 & 353 & 0.006 & 374 & 0.008 \\
4   & 19 & 286 & 0.006 & 4   & 0.008 \\
468 & 19 & 295 & 0.006 & 429 & 0.008 \\
443 & 19 & 323 & 0.006 & 441 & 0.008 \\
497 & 19 & 223 & 0.006 & 498 & 0.007 \\
398 & 19 & 196 & 0.006 & 439 & 0.007 \\
436 & 19 & 340 & 0.005 & 357 & 0.007 \\
\hline
\end{tabular}
\label{tab:hubs_mcfm_s}
\end{table}

\begin{table}[ht]
\centering
\caption{Top 20 nodes by degree, HITS hub scores, and HITS authority scores in the INEP after 2000 added links ($N=500$).}
\begin{tabular}{|c|c||c|c||c|c|}
\hline
\textbf{Node} & \textbf{Degree} & \textbf{Node} & \textbf{Hub Score} & \textbf{Node} & \textbf{Authority Score} \\
\hline
3 & 90 & 2 & 0.027 & 6 & 0.020 \\
2 & 90 & 3 & 0.026 & 7 & 0.018 \\
0 & 85 & 1 & 0.025 & 5 & 0.016 \\
1 & 82 & 0 & 0.022 & 4 & 0.014 \\
5 & 76 & 108 & 0.008 & 8 & 0.012 \\
7 & 69 & 21 & 0.008 & 459 & 0.010 \\
6 & 68 & 98 & 0.008 & 472 & 0.010 \\
4 & 66 & 430 & 0.008 & 267 & 0.010 \\
472 & 43 & 54 & 0.007 & 406 & 0.009 \\
459 & 43 & 196 & 0.007 & 9 & 0.009 \\
\hline
\end{tabular}
\label{tab:hubs_cem}
\end{table}

\newpage


\end{document}